\documentclass[10pt,journal]{IEEEtran}

\usepackage{amsmath,graphics,amssymb,epsfig, color,cite,bm}
\usepackage{array}
\usepackage{multirow}
\usepackage{enumerate}
\usepackage{algorithm}
\usepackage{algpseudocode}
\usepackage{algcompatible}
\usepackage{tablefootnote}
\usepackage[flushleft]{threeparttable}
\usepackage{booktabs}
\usepackage[right=0.625in, left=0.625in, top=0.7in, bottom=1.1in]{geometry}

\algblockdefx{FORP}{ENDFORP}[1]%
  {\textbf{for }#1 \textbf{do in parallel}}%
  {\textbf{end}}
\algblockdefx{FOR}{ENDFOR}[1]%
  {\textbf{for }#1 \textbf{do}}%
  {\textbf{end}}
\algblockdefx{IF}{ENDIF}[1]%
  {\textbf{if }#1 \textbf{then}}%
  {\textbf{end}}
\algblockdefx{PROC}{ENDPROC}[1]
    {\textbf{procedure }#1 {}}
    {\textbf{end procedure}}
  
\algnewcommand\algorithmicprocedure{\textbf{function}}
\algnewcommand\FUNC{\item[\algorithmicprocedure]}%
\algnewcommand\algorithmicendprocedure{\textbf{end function}}
\algnewcommand\ENDFUNC{\item[\algorithmicendprocedure]}%

\usepackage{float}
\usepackage{makecell}
\usepackage[caption=false,font=footnotesize]{subfig}

\makeatletter

\newcommand{\vast}{\bBigg@{4.5}}
\newcommand{\Vast}{\bBigg@{7.5}}

\makeatother

\begin{document}
    \title{Anti-Jamming Modulation for OFDM Systems \\ under Jamming Attacks}
    
    \author{Jaewon Yun, Joohyuk Park, and Yo-Seb Jeon
        \thanks{Jaewon Yun, Joohyuk Park, and Yo-Seb Jeon are with the Department of Electrical Engineering, POSTECH, Pohang, Gyeongbuk 37673, South Korea (e-mail: jaewon.yun@postech.ac.kr; joohyuk.park@postech.ac.kr; yoseb.jeon@postech.ac.kr).}
        % \thanks{This work was supported in part by Korea Research Institute for defense Technology planning and advancement(KRIT) - Grant funded by the Defense Acquisition Program Administration(DAPA) (KRIT-CT-22-078) and in part by the National Research Foundation of Korea(NRF) grant funded by the Korea government(MSIT) (No. RS-2024-00453301).}
    }
    \vspace{-2mm}	
    
    \maketitle
    \vspace{-12mm}
    
    \begin{abstract} %199단어
         Orthogonal frequency division multiplexing (OFDM) systems are inherently vulnerable to jamming attacks due to the independent transmission of data symbols across subcarriers.
         In this paper, we propose a novel anti-jamming OFDM scheme to ensure robust communication even under severe jamming attacks while maintaining high spectral efficiency.
         The core idea is to utilize a spreading matrix that transforms a data symbol vector into a higher-dimensional modulated vector, thereby exploiting both the spreading gain and the frequency diversity gain to mitigate jamming attacks.
         To recover the transmitted data symbols, we develop an efficient maximum likelihood detection (MLD) method that achieves optimal detection performance with significantly reduced computational complexity.
         Furthermore, we derive the theoretical bit error rate (BER) upper bound and the optimal modulation order that minimizes the BER while preserving spectral efficiency according to the jamming environment.
         To address practical scenarios where jamming attacks are unknown and dynamic, we establish a jamming-adaptive communication framework.
         This framework enables the system to estimate the jamming parameters and adapt to the dynamic environment with the optimal modulation order.
         Simulation results demonstrate that the proposed scheme significantly outperforms existing OFDM schemes in both BER and effective throughput, validating its robustness under various and dynamic jamming scenarios.
    \end{abstract}
    
    \begin{IEEEkeywords} 
       Anti-jamming modulation, anti-jamming orthogonal frequency-division multiplexing, jamming attacks, jamming parameter estimation, jamming robust communications 
    \end{IEEEkeywords}

    \section{Introduction}\label{Sec:Intro}

    Orthogonal frequency division multiplexing (OFDM) has been widely adopted in modern wireless communication systems due to its high spectral efficiency and robustness against multipath fading.
    However, due to the independent transmission of data symbols across subcarriers, OFDM systems are inherently vulnerable to external interference or malicious jamming attacks \cite{Challenge1,Challenge2,Challenge3}.
    Even with advanced error-correcting codes, reliable communication becomes unachievable if the errors caused by jamming attacks exceed the correction capability or if critical resources are corrupted.
    Consequently, such vulnerability poses a severe threat to safety-critical applications, including military communications, autonomous systems, and infrastructure networks, where communication breakdown can lead to catastrophic consequences.

    In practice, jamming attacks often manifest as active interference in various scenarios, including broadband barrage jamming, frequency-selective jamming (e.g., partial-band and sweep jamming), time-selective pulse jamming, and unpredictable jamming (e.g., random and smart jamming) \cite{RJ, SmartJammer,Survey2}.
    The most widely adopted approaches to mitigate these threats are frequency-hopping spread spectrum (FH-SS) and direct-sequence spread spectrum (DS-SS) \cite{FH, DS-SS,Survey1}.
    However, these techniques are typically designed for single-carrier waveforms, often resulting in structural mismatches or reduced efficiency when integrated into modern OFDM systems.
    Beyond these classical approaches, a variety of advanced countermeasures have been investigated in \cite{RNN, AAJ, DPC, Crypto, AJ1, AJ2}, including recurrent neural networks (RNNs) that predict jamming patterns \cite{RNN}, active anti-jamming (AAJ) utilizing the jamming signal for transmission \cite{AAJ}, and dirty paper coding (DPC) for interference pre-cancellation \cite{DPC}.
    Nevertheless, the integration of these techniques into OFDM systems remains largely unexplored, and most of them rely on prior knowledge of jamming characteristics or interference states, an assumption that is difficult to guarantee in practical hostile environments.

    To address these limitations, the integration of classical anti-jamming techniques into OFDM systems has been investigated in \cite{FH2, FH3, MC1, MC2, OFDMJAM1, OFDMJAM2, OFDMJAM3}.
    For example, frequency hopping OFDM (FH-OFDM), which adopts the principle of FH-SS, has been studied in \cite{FH2, FH3}.
    This technique mitigates interference by hopping active subcarriers based on pseudorandom hopping patterns, thereby evading frequency-selective jamming.
    However, FH-OFDM inherently limits spectral efficiency, as only a subset of the available subcarriers is activated for data transmission within each OFDM symbol.
    Similarly, multi-carrier code division multiple access (MC-CDMA), which follows the principle of DS-SS, has been developed in \cite{MC1, MC2}. 
    This technique enhances robustness by spreading each data symbol across multiple subcarriers in the frequency domain using spreading sequences.
    Although such spreading provides robustness through processing gain, the achievable spectral efficiency is inversely proportional to the spreading factor.
    Consequently, achieving sufficient robustness in these schemes leads to severe spectral efficiency degradation, which fundamentally contradicts the primary objective of employing OFDM.

    Anti-jamming OFDM techniques beyond classical FH-SS or DS-SS approaches have been developed in \cite{WHT1, WHT2, IM:Jamming1, IM:Jamming2}.
    A representative example is precoded OFDM, such as Walsh-Hadamard transform (WHT)-OFDM, which employs precoding matrices to exploit the spreading gain and frequency diversity gain to mitigate jamming attacks \cite{WHT1, WHT2}.
    However, these schemes predominantly rely on linear detectors, such as minimum mean square error (MMSE), due to the high computational complexity associated with large precoding matrices, leading to an unavoidable error floor under severe jamming attacks.
    Another example is OFDM with index modulation (OFDM-IM) \cite{IM}, which conveys additional information bits through the indices of active subcarriers, providing inherent jamming robustness via its sparse activation pattern \cite{IM:Jamming1, IM:Jamming2}.
    Nevertheless, its passive interference avoidance nature renders it vulnerable when active subcarriers are corrupted by severe jamming attacks \cite{ICTC1,ICTC2}.
    Although this vulnerability can be partially alleviated by enabling intelligent subcarrier selection through reinforcement learning (RL) \cite{RL}, this approach often requires significant convergence time, limiting its ability to adapt to unpredictable and rapidly varying jamming environments.
    Therefore, developing an anti-jamming OFDM scheme that simultaneously achieves high spectral efficiency and enhanced robustness against unpredictable and severe jamming attacks remains a critical open challenge.

    In this paper, our major goal is to develop an anti-jamming OFDM scheme that maintains high spectral efficiency while ensuring robust communication under various and dynamic jamming scenarios without prior knowledge of jamming parameters.
    To achieve this goal, we first propose novel anti-jamming modulation and demodulation methods.
    For the proposed methods, we derive the theoretical bit error rate (BER) upper bound to determine the optimal modulation order according to the jamming environment.
    To ensure practicality, we extend the proposed scheme into a jamming-adaptive communication framework capable of adapting to dynamic jamming scenarios.
    Through simulations, we demonstrate that the proposed scheme significantly outperforms existing OFDM schemes in both BER and effective throughput.
    The major contributions of this paper are summarized as follows:
    \begin{itemize}
        \item 
        We propose a novel anti-jamming modulation method designed to enhance jamming robustness while preserving the spectral efficiency for OFDM systems.
        The core idea is to utilize a spreading matrix that transforms a quadrature amplitude modulation (QAM) symbol vector into a higher-dimensional modulated vector.
        This ensures that each QAM symbol is transmitted across multiple subcarriers instead of being mapped to a single subcarrier.
        As a result, even if some subcarriers are heavily jammed, the QAM symbols can still be successfully demodulated by exploiting the unjammed subcarriers.
        To address the challenge where the indices of the jammed subcarriers are unknown, we develop an exhaustive maximum likelihood detection (MLD) method.
        This method jointly estimates the QAM symbol vector and the indices of the jammed subcarriers, assuming that the jamming variance is known.
        We also devise an efficient MLD method via the rearrangement inequality, which reduces computational burden without compromising detection performance.
        
        \item
        We derive the theoretical BER upper bound of the proposed anti-jamming modulation and demodulation methods.
        This upper bound is expressed as a function of the modulation order and the channel and jamming environments.
        Based on this analytical result, we formulate an optimization problem that determines the optimal modulation order by using the derived BER upper bound as the objective function, thereby minimizing the BER while preserving spectral efficiency, i.e., maximizing the effective throughput.
        
        \item
        We propose a jamming-adaptive communication framework to enable robust communication under dynamic jamming scenarios without prior knowledge of jamming parameters.
        This framework comprises two phases: (i) a jamming estimation phase and (ii) a jamming adaptation phase.
        In the jamming estimation phase, we devise an approximate MLD method based on the generalized likelihood ratio test (GLRT) principle to estimate the jamming variance and the average number of jammed subcarriers.
        In the jamming adaptation phase, we utilize the optimal modulation order based on the estimated parameters and employ the efficient MLD method to recover the symbols.
        
        \item
        Simulation results demonstrate that the proposed scheme significantly outperforms existing OFDM schemes in both BER and effective throughput under various jamming scenarios.
        The proposed scheme also remains robust against practical impairments, such as channel estimation errors, confirming its superiority even in realistic scenarios.
        Furthermore, without any prior knowledge of the jamming parameters, the proposed jamming-adaptive communication framework achieves a BER more than 20 times lower than that of the existing schemes under random jamming at a signal-to-jamming ratio of $-20$~dB.
        As a result, it improves the effective throughput by more than 10\% over the existing schemes at the same spectral efficiency.
        %Furthermore, without any prior knowledge of the jamming parameters, the proposed jamming-adaptive communication framework achieves a BER of approximately $5 \times 10^{-3}$ under severe jamming, even at twice the spectral efficiency of the existing schemes, which remain above $10^{-1}$.
        %As a result, the proposed framework maintains the target throughput of 1 and 0.5~bps/Hz regardless of the jamming scenario, whereas the existing schemes suffer from throughput degradation.}
        The proposed framework also adapts to dynamic jamming scenarios by estimating the jamming parameters and continuously updating the optimal modulation order.
    \end{itemize}

    \begin{figure*}
        \centering
        \includegraphics[width=0.98\textwidth]{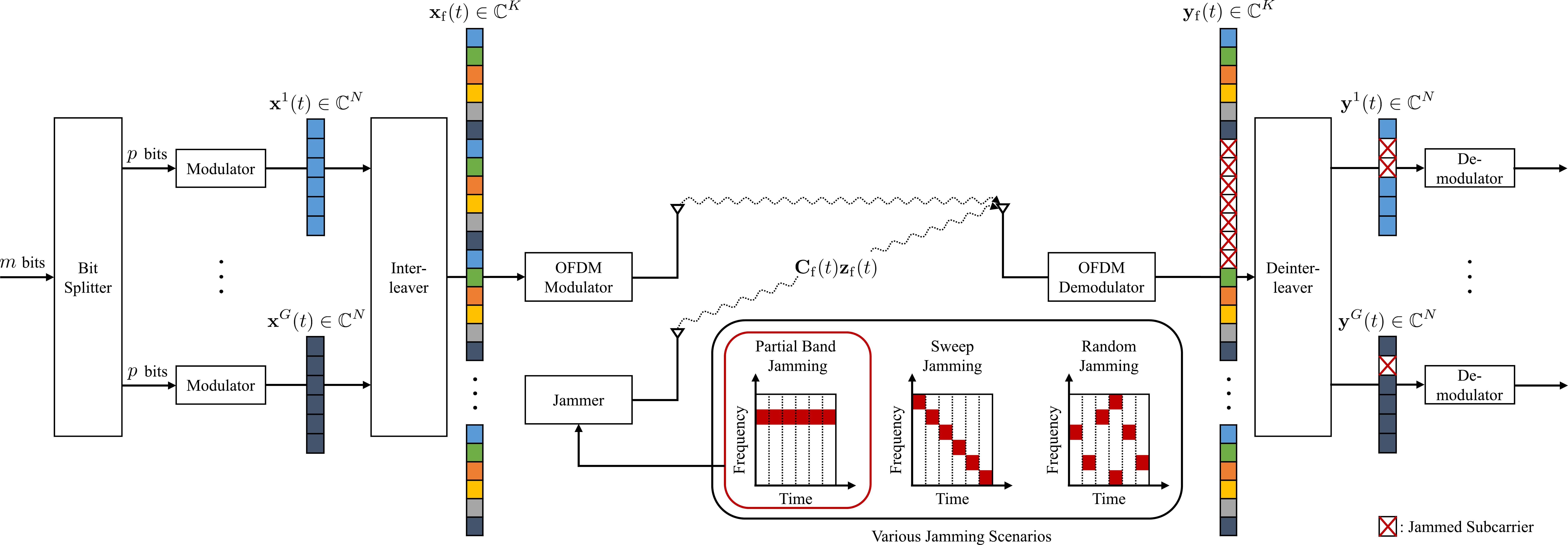}
        \vspace{-2mm}
        \caption{Illustration of a point-to-point OFDM communication system under various jamming scenarios when employing interleaving and deinterleaving processes at the transmitter and receiver, respectively.}
        \vspace{-3mm}
        \label{Fig_1}
    \end{figure*}

    {\em Notation:} 
    Upper-case and lower-case boldface letters denote matrices and column vectors, respectively.
    $\mathbb{E}[\cdot]$ denotes the statistical expectation, and $(\cdot)^{\sf T}$ and $(\cdot)^{\sf H}$ represent the transpose and Hermitian, respectively.
    $|a|$ and $|{\mathcal A}|$ denote the magnitude of a complex scalar $a$ and the cardinality of a set ${\mathcal A}$, respectively.
    $a_i$ represents the $i$-th element of a vector ${\bf a}$.
    $\|{\bf a}\| = \sqrt{{\bf a}^{\sf H}{\bf a}}$ is the Euclidean norm of a complex vector ${\bf a}$.
    ${\sf diag}({\bf a})$ denotes a diagonal matrix with the elements of vector ${\bf a}$ on its main diagonal.
    $\mathcal{CN}(\bm{\mu},{\bf \Sigma})$ represents the circularly-symmetric complex Gaussian distribution with mean vector $\bm{\mu}$ and covariance matrix ${\bf \Sigma}$.
    ${\bf 0}_N$ is an $N$-dimensional zero vector.
    ${\bf I}_N$ is an $N \times N$ identity matrix.

    \section{System Model}\label{Sec:System}

    Consider a point-to-point OFDM communication system under jamming attacks, as illustrated in Fig.~\ref{Fig_1}.
    As shown in Fig.~\ref{Fig_1}, $m$ information bits are transmitted across $K$ subcarriers in each OFDM symbol, where $K$ denotes the total number of subcarriers.
    These bits are assumed to be divided equally into $G$ blocks, with each block containing $p = m/G$ bits.
    For each block, the $p$ bits are jointly modulated into an $N$-dimensional modulated vector.
    Note that a conventional OFDM scheme using $M$-QAM corresponds to the special case where each block contains a single $M$-QAM symbol, i.e., $p=\log_2 M$ and $N=1$.
    A {\em per-block} modulated vector associated with block $g$ of the $t$-th OFDM symbol is denoted as
    \begin{align}\label{eq:def_symbol}
        {\bf x}^{g}(t)= [x_1^{g}(t), \dots, x_N^{g}(t)]^{\sf T},~~\forall g\in\{1,\ldots,G\},
    \end{align}
    where $x_n^{g}(t)$ is the $n$-th entry of ${\bf x}^{g}(t)$.
    Before mapping the $G$ modulated vectors into $K$ subcarriers, a frequency interleaver is applied to avoid burst errors\footnote{Jamming attacks often target contiguous subcarriers within a specific bandwidth.
    In this case, burst errors can occur for the entries of the modulated vector transmitted across those contiguous subcarriers.} due to jamming attacks.
    Fig.~\ref{Fig_1} illustrates how interleaving redistributes the entries of the modulated vectors across subcarriers to mitigate burst errors caused by various jamming scenarios.
    As a result, the frequency-domain transmitted vector of the $t$-th OFDM symbol is given by
    \begin{align}\label{eq:interleaver}
        {\bf x}_{\rm f}(t)= {\sf Interleaver}({\bf x}^{1}(t), {\bf x}^{2}(t), \dots, {\bf x}^{G}(t)).
    \end{align}
    This vector is assumed to be normalized to satisfy a power constraint of $\mathbb{E}[\|{\bf x}_{\rm f}(t)\|^2] = K$.
    The subsequent processing steps follow those of a conventional OFDM transmission process.
    The time-domain transmitted vector is obtained by applying the $K$-point inverse fast Fourier transform (IFFT) to the frequency-domain transmitted vector ${\bf x}_{\rm f}(t)$, i.e.,
    \begin{align}
        {\bf x}_{\rm t}(t) = \frac{1}{\sqrt{K}}{\bf W}_K^{\sf H}{\bf x}_{\rm f}(t),
    \end{align}
    where ${\bf W}_K$ denotes the $K$-point discrete Fourier transform (DFT) matrix such that ${\bf W}_K^{\sf H}{\bf W}_K = K{\bf I}_K$.
    Finally, the time-domain transmitted vector ${\bf x}_{\rm t}(t)$ is processed for transmission by appending a cyclic prefix (CP), followed by parallel-to-serial (P/S) conversion and digital-to-analog conversion.

    In this work, we focus on communications under jamming attacks and therefore assume the presence of a jamming signal that interferes with the frequency-domain transmitted vector across $K$ subcarriers.
    In practice, jamming attacks manifest in various scenarios, such as partial-band jamming, sweep jamming, and random jamming \cite{RJ, SmartJammer, Survey2}.
    To provide a unified model for various jamming scenarios, we introduce a frequency-domain jamming indicator vector ${\bf c}_{\rm f}(t)\in\{0,1\}^K$ whose $k$-th entry indicates whether the jamming signal is present at subcarrier $k$ of the $t$-th OFDM symbol.
    Note that ${\bf c}_{\rm f}(t)$ can vary over time in certain jamming scenarios (e.g., sweep jamming, random jamming, and pulse jamming).
    Utilizing this vector, we model the frequency-domain jamming vector of the $t$-th OFDM symbol as
    \begin{align}
        {\bf C}_{\rm f}(t){\bf z}_{\rm f}(t) = {\sf diag}({\bf c}_{\rm f}(t)) {\bf z}_{\rm f}(t),
    \end{align}
    where ${\bf z}_{\rm f}(t)$ is the frequency-domain jamming symbol vector.
    We assume that the entries of ${\bf z}_{\rm f}(t)$ are independent and follow a circularly symmetric complex Gaussian distribution with variance $\sigma_z^2$ (i.e., ${\bf z}_{\rm f}(t)\sim \mathcal{CN}({\bf 0}_K,\sigma_z^2{\bf I}_K)$).

    At the receiver, the CP is first removed from the received signal, followed by the $K$-point FFT to obtain the frequency-domain received vector denoted as ${\bf y}_{\rm f}(t)$.
    In the presence of the jamming signal, the frequency-domain received vector of the $t$-th OFDM symbol is expressed as
    \begin{align}\label{eq:system}
        {\bf y}_{\rm f}(t) = {\bf H}_{\rm f}(t){\bf x}_{\rm f}(t) + {\bf C}_{\rm f}(t){\bf z}_{\rm f}(t) + {\bf w}_{\rm f}(t),
    \end{align}
    where ${\bf H}_{\rm f}(t) = {\sf diag}({\bf h}_{\rm f}(t))$, ${\bf h}_{\rm f}(t) = [h_{{\rm f},1}(t), \dots, h_{{\rm f},K}(t)]^{\sf T}$, $h_{{\rm f},k}(t)$ is the channel frequency response (CFR) at subcarrier $k$, and ${\bf w}_{\rm f}(t)$ is the frequency-domain additive white Gaussian noise (AWGN) vector with ${\bf w}_{\rm f}(t) \sim \mathcal{CN}({\bf 0}_K,\sigma_w^2{\bf I}_K)$.
    To focus on ideal receiver operation, it is assumed that the receiver has perfect knowledge of the channel matrix ${\bf H}_{\rm f}(t)$ and the noise variance $\sigma_w^2$.
    The impacts of channel estimation errors and time-varying channels are also evaluated in Figs.~\ref{Fig_6}(a) and \ref{Fig_6}(b), respectively.
    The received vector associated with each per-block modulated vector in \eqref{eq:def_symbol} can be reconstructed by deinterleaving\footnote{We assume that the interleaving pattern applied at the transmitter is known at the receiver.} the frequency-domain received vector ${\bf y}_{\rm f}(t)$, as illustrated in Fig.~\ref{Fig_1}.
    As a result, the per-block received vector associated with ${\bf x}^g(t)$ is expressed as
    \begin{align}\label{eq:blocksystem}
        {\bf y}^g(t) &= {\sf diag}({\bf h}^g(t)){\bf x}^g(t) + {\sf diag}({\bf c}^g(t)) {\bf z}^g(t) + {\bf w}^g(t) \nonumber \\
        &= {\bf H}^g(t){\bf x}^g(t) + {\bf C}^g(t){\bf z}^g(t) + {\bf w}^g(t),
    \end{align}
    where ${\bf H}^g(t)$, ${\bf C}^g(t){\bf z}^g(t)$, and ${\bf w}^g(t)$ are the channel matrix, the jamming vector, and the noise vector associated with block $g$ of the $t$-th OFDM symbol, respectively.
    Note that the jamming indicator vector ${\bf c}^g(t)\in\{0,1\}^N$ can vary across blocks and OFDM symbols depending on both the jamming scenarios and the deinterleaving patterns as shown in Fig.~\ref{Fig_1}.

    \begin{figure}[t]
        \centering
        \includegraphics[width=0.98\columnwidth]{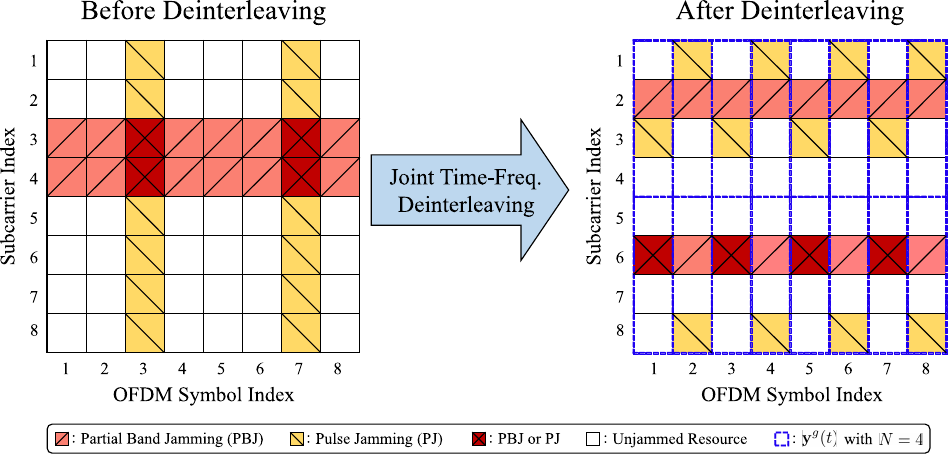}
        \vspace{-2mm}
        \caption{Illustration of the effectiveness of joint time-frequency interleaving against partial-band jamming (PBJ) and pulse jamming (PJ).
        A simplified example with $K=8$ subcarriers and $T_{\rm int}=8$ OFDM symbols is depicted.}
        \vspace{-3mm}
        \label{Fig_2}
    \end{figure}

    \textbf{Remark 1 (Joint Time-Frequency Interleaving):} 
    The frequency interleaver in \eqref{eq:interleaver} effectively distributes frequency-selective jamming (e.g., partial-band jamming) but remains vulnerable to pulse jamming, which corrupts all subcarriers within a specific OFDM symbol.
    To address this, the interleaver can be extended to joint time-frequency interleaving over $T_{\rm int}$ OFDM symbols with the input set $\{{\bf x}^g(t)\}_{g=1, t=1}^{G, T_{\rm int}}$, redistributing the entries of the modulated vectors across both the time and frequency domains.
    As shown in Fig.~\ref{Fig_2}, this process disperses burst errors under either partial-band jamming or pulse jamming into sparse errors within each per-block received vector ${\bf y}^g(t)$.
    It is worth noting that this extension introduces a processing delay proportional to $T_{\rm int}$.

    \section{Proposed Anti-Jamming Modulation and Demodulation Methods}\label{Sec:AJM}

    In this section, we propose a novel anti-jamming modulation method designed to enhance jamming robustness while preserving the spectral efficiency for OFDM systems.
    We then develop two demodulation methods, referred to as {\em exhaustive MLD} and {\em efficient MLD}, tailored to support our anti-jamming modulation method.
    Our modulation and demodulation methods operate on a block-by-block basis.
    Therefore, in the rest of this section, we focus on describing the operation for an individual block while omitting the block and time indices $g$ and $t$ for brevity.
    
    \subsection{Proposed Anti-Jamming Modulation}\label{Sec:Modulation}
    The core idea of our anti-jamming modulation method is to utilize a spreading matrix that transforms a QAM symbol vector into a higher-dimensional modulated vector.
    This ensures that each QAM symbol is transmitted across multiple subcarriers instead of being mapped to a single subcarrier.
    As a result, even if some subcarriers are heavily jammed, the QAM symbols can still be successfully demodulated by exploiting the unjammed subcarriers.

    To implement this idea, the proposed modulation method first maps the $p$ bits to $S$ symbols using $M$-QAM with a modulation order $M=2^{p/S}$.
    Let ${\bf s} \in {\mathcal X}_M^S$ be the resulting symbol vector, where ${\mathcal X}_M$ is the normalized constellation set of $M$-QAM such that $\mathbb{E}[\|{\bf s}\|^2]=S$.
    Then, the symbol vector ${\bf s}$ is transformed into an $N$-dimensional modulated vector by multiplying a spreading matrix ${\bf U} \in \mathbb{C}^{N\times S}$ as follows:
    \begin{align}
        {\bf x} = {\bf U}{\bf s} = [{\bf u}_1^{\sf H}{\bf s}, \dots, {\bf u}_N^{\sf H}{\bf s}]^{\sf T},
    \end{align}
    where ${\bf U} = [{\bf u}_1,\cdots,{\bf u}_N]^{\sf H}$.
    To construct the spreading matrix for the modulation order $M$, the symbol vector length is determined as $S = p/\log_2 M$.
    Then, a submatrix ${\bf U}_0^\prime$ is formed by selecting the first $S$ columns of a predetermined random unitary matrix ${\bf U}_0$.
    This submatrix is scaled by $\sqrt{N/S}$ to ensure that the power of the modulated vector ${\bf x}$ is preserved as $N$.
    Consequently, our spreading matrix is expressed as
    \begin{align}\label{eq:unitary}
        {\bf U} = \sqrt{\frac{N}{S}} {\bf U}_0^\prime.
    \end{align}
    Note that the above construction ensures that the power of the modulated vector is given by
    \begin{align}
        {\mathbb E}[\|{\bf x}\|^2] = \frac{N}{S} \cdot {\mathbb E}[\|{\bf s}\|^2] = \frac{N}{S} \cdot S = N.
    \end{align}
    This result and the transmission of $p$ bits across $N$ subcarriers confirm that our modulation method strictly preserves both the spectral efficiency and average transmit power of a conventional OFDM scheme, regardless of the modulation order $M$.
    The unitary matrix ${\bf U}_0$ is assumed to be fixed and known to both the transmitter and the receiver, enabling the generation of identical spreading matrices for different modulation orders $M$.
    
    A key advantage of the proposed modulation method is that it utilizes a spreading mechanism to achieve a spreading gain by distributing the information of the $S$ QAM symbols across $N$ subcarriers.
    Specifically, even in the presence of infinite jamming power on $J$ subcarriers (i.e., $J = \|{\bf c}\|_0$), the proposed demodulation method detailed in Sec.~\ref{Sec:FullMLD} effectively discards these subcarriers as they provide no valid information.
    Consequently, from the remaining $N-J$ unjammed subcarriers, the $S$ symbols can be perfectly recovered in the absence of noise, provided that the linear algebraic solvability condition $N-J \ge S$ is satisfied.
    Accordingly, to satisfy this condition, a sufficient spreading gain of $N/S$ is required, for which our modulation method utilizes a higher modulation order $M = 2^{p/S}$ compared to the conventional modulation order of $M=2^{p/N}$.
    However, utilizing a higher-order QAM constellation inevitably degrades the performance due to the reduced minimum Euclidean distance.
    Therefore, it is essential to optimize the modulation order $M$ to satisfy the solvability condition while minimizing detection errors, which will be discussed in Sec.~\ref{Sec:Optimization}.
    It is worth noting that even in barrage jamming or jamming-free scenarios, the spreading mechanism provides a spreading gain of $N/S$, and a frequency diversity gain against deep fading.

    \textbf{Remark 2 (Additional Robustness against Smart Jamming):}
    Beyond the spreading gain and the frequency diversity gain, the spreading matrix ${\bf U}$ provides additional robustness against smart jamming attacks that adapt to the observed transmission, such as reactive and correlated jamming.
    Since ${\bf U}$ distributes the information across all $N$ subcarriers, all the subcarriers are always active, and thus it is difficult for the reactive jammer to determine which subcarriers to attack \cite{RL}.
    Moreover, without knowledge of ${\bf U}$, it is difficult for the correlated jammer to infer the structure of the modulated vector ${\bf x}={\bf U}{\bf s}$.
    Even when the jamming signal is generated based on the overheard transmission, it passes through the random channel between the jammer and the receiver, and thus manifests as additional random interference at the receiver.

    \subsection{Exhaustive MLD Method with Known Jamming Variance}\label{Sec:FullMLD}
    We now devise the exhaustive MLD method for the proposed anti-jamming modulation method, applicable when the jamming variance is known\footnote{A more practical demodulation method applicable for unknown jamming variance will be introduced in Sec.~\ref{Sec:ApproxMLD}.}.
    Unlike conventional MLD which performs symbol-by-symbol detection, the proposed MLD jointly detects the symbol vector ${\bf s}$ and the jamming indicator vector ${\bf c}$ from the received vector ${\bf y}$ by maximizing the likelihood function.
    This is achieved by comparing the likelihood functions of ${\bf y}$ for all combinations of the symbol vector ${\bf s}$ and the jamming indicator vector ${\bf c}$.
    The exhaustive MLD method is formulated as the following \textit{joint optimization problem}:
    \begin{align}\label{eq:ML}
        (\hat{\bf s}_{{\rm ML}}, \hat{\bf c}_{\rm ML}) = \underset{{\bf s} \in {\mathcal X}_M^S, {\bf c} \in \{0,1\}^N}{\arg\!\max}~p({\bf y}|{\bf s},{\bf c}),
    \end{align}
    where $p({\bf y}|{\bf s},{\bf c})$ is the likelihood function of the received vector ${\bf y}$ given the symbol vector ${\bf s}$ and the jamming indicator vector ${\bf c}$.
    Recall that the received vector ${\bf y}$ in the presence of the jamming signal is given by
    \begin{align}\label{eq:systemmodel}
        {\bf y} = {\bf H}{\bf x} + {\bf C}{\bf z} + {\bf w}.
    \end{align}
    Based on the distributions of the jamming signal and the noise, the likelihood function $p({\bf y}|{\bf s},{\bf c})$ is expressed as
    \begin{align}
        p({\bf y}|{\bf s},{\bf c})
        =& \prod_{i=1}^N \frac{1}{\pi(c_i\sigma_z^2 + \sigma_w^2)} \exp\left(- \frac{|y_i - h_i {\bf u}_i^{\sf H}{\bf s}|^2}{c_i\sigma_z^2 + \sigma_w^2} \right),
    \end{align}
    which simplifies to
    \begin{align}\label{eq:LF_full}
        p({\bf y}|{\bf s},{\bf c}) =&
        \frac{1}{\pi^{N}(\sigma_z^2 + \sigma_w^2)^{\|{\bf c}\|_0} (\sigma_w^2)^{N-\|{\bf c}\|_0}} \nonumber \\
        &\times \exp\left(- \sum_{i=1}^N \frac{|y_i - h_i {\bf u}_i^{\sf H}{\bf s}|^2}{c_i\sigma_z^2 + \sigma_w^2} \right).
    \end{align}
    Consequently, the joint optimization problem of the exhaustive MLD method is given by
    \begin{align}\label{eq:MLD}
        (\hat{\bf s},\hat{\bf c})
        &= \underset{{\bf s}\in\mathcal{X}_M^S, {\bf c}\in\{0,1\}^N}{\arg\!\max}~
        \frac{1}{(\sigma_z^2/ \sigma_w^2 + 1)^{\|{\bf c}\|_0}} \nonumber \\
        &~~~~~~~~~~~~~~\qquad \times \exp\left(- \sum_{i=1}^N \frac{|y_i - h_i {\bf u}_i^{\sf H}{\bf s}|^2}{c_i\sigma_z^2 + \sigma_w^2} \right),
    \end{align}
    where terms irrelevant to the comparison are omitted.
    
    The expression in \eqref{eq:MLD} shows that the observation from the $i$-th subcarrier contributes by calculating the weighted squared Euclidean distance to the ideal received signal $h_i{\bf u}_i^{\sf H}{\bf s}$, where ${\bf u}_i^{\sf H}$ denotes the $i$-th row vector of the spreading matrix ${\bf U}$.
    This indicates that the detection of the symbol vector ${\bf s}$ at the receiver jointly leverages all $N$ observations from the $N$ subcarriers, which is a distinctive feature of our anti-jamming modulation method.
    Specifically, when the $i$-th subcarrier is jammed (i.e., $c_i = 1$), the jamming variance is added to the noise variance, thereby assigning a lower weight to the corresponding observation in the MLD method.
    Conversely, when the $i$-th subcarrier is unjammed (i.e., $c_i = 0$), the observation is weighted by the inverse of the noise variance, thereby assigning it a relatively higher weight.
    To guarantee the reliable recovery of the symbol vector ${\bf s}$ from these highly weighted observations even under severe jamming attacks, the condition $N-J \ge S$ is necessary.
    This implies that even in an extreme scenario where only a single subcarrier is left unjammed (i.e., $N-J=1$), the above condition is satisfied by utilizing the highest modulation order (i.e., with $S=1$).
    Consequently, the exhaustive MLD method recovers the symbol vector ${\bf s}$ by aggregating the energy of each QAM symbol distributed by the spreading matrix ${\bf U}$, while jointly identifying the jamming indicator vector ${\bf c}$.
    
    We now analyze the computational complexity of the exhaustive MLD method described above.
    As shown in \eqref{eq:MLD}, the exhaustive MLD method requires an exhaustive search over all possible combinations of the symbol vector ${\bf s}$ and the jamming indicator vector ${\bf c}$.
    Given that ${\bf s}$ is selected from the $M$-QAM constellation set ${\mathcal X}_M^S$ of size $M^S$, and ${\bf c}$ is selected from the binary vector set $\{0,1\}^N$ of size $2^N$, the total number of hypotheses corresponds to $M^S \cdot 2^N$.
    It is particularly noteworthy that, since the number of transmitted bits is fixed to $p$ (i.e., $M^S = 2^p$), the size of the symbol search space remains constant regardless of the modulation order $M$.
    Since the likelihood calculation for each hypothesis involves operations over $N$ subcarriers, the computational complexity is given by $\mathcal{O}(M^S 2^N N)$.
    This analysis highlights that, despite the constant search space for the symbol vector, the overall complexity of the exhaustive MLD method becomes prohibitive in practical systems due to the exponential dependence on $N$, motivating the development of a low-complexity alternative.
    
    \subsection{Efficient MLD Method with Known Jamming Variance}\label{Sec:LowcompMLD}
    
    To address the prohibitive computational complexity of the exhaustive MLD method discussed in Sec.~\ref{Sec:FullMLD}, we develop an efficient MLD method that solves the joint optimization problem without an exhaustive search.
    Instead of searching over all possible jamming indicator vectors ${\bf c} \in \{0,1\}^N$ jointly with the symbol vector, we exploit the mathematical structure of the likelihood function to decompose the problem.
    
    Let ${\bf e}({\bf s})$ denote the residual error vector associated with a candidate symbol vector ${\bf s}$, defined as
    \begin{align}\label{eq:residual}
        {\bf e}({\bf s}) = {\bf y} - {\bf H} {\bf U} {\bf s}.
    \end{align}
    We observe from the likelihood function in \eqref{eq:LF_full} that for a fixed symbol vector ${\bf s}$ and a fixed number of jammed subcarriers $J = \|{\bf c}\|_0$, maximizing the likelihood is equivalent to minimizing the weighted sum of squared residual errors.
    Therefore, for a fixed $J$, the optimal jamming indicator vector $\hat{\bf c}^{(J)}$ is determined by
    \begin{align}
        \hat{\bf c}^{(J)} 
        &= \underset{{\bf c}:\|{\bf c}\|_0=J}{\arg\!\min} \sum_{i=1}^N \frac{|e_i({\bf s})|^2}{c_i\sigma_z^2 + \sigma_w^2}, \end{align}
    which leads to
    \begin{align}\label{eq:sort_c}
        \hat{c}_i^{(J)} = \begin{cases} 1, & |e_i({\bf s})| \ge |{\sf sort}_J({\bf e}({\bf s}))|, \\ 0, & \text{otherwise},
        \end{cases}
    \end{align}
    where ${\sf sort}_i({\bf a})$ denotes the $i$-th largest entry of a vector ${\bf a}$ in terms of magnitude.
    According to the rearrangement inequality, the objective function above is minimized when the largest denominators are assigned to the terms with the largest residual magnitudes.
    Since the jamming-plus-noise variance is strictly larger than the noise variance (i.e., $\sigma_z^2 + \sigma_w^2 > \sigma_w^2$), the optimal jamming indicator vector $\hat{\bf c}^{(J)}$ must assign $c_i=1$ to the indices corresponding to the $J$ largest magnitudes of the residual error vector ${\bf e}({\bf s})$.
    This implies that for any candidate symbol vector ${\bf s}$, the indices of the jammed subcarriers are uniquely determined by simply sorting the magnitudes of the residual errors.

    By substituting this optimal jamming indicator vector $\hat{\bf c}^{(J)}$ back into \eqref{eq:LF_full}, we obtain the \textit{profile likelihood function} (i.e., the likelihood evaluated at the optimal $\hat{\bf c}^{(J)}$) for a given ${\bf s}$ and $J$ as \eqref{eq:LF_profile}.
    \begin{figure*}
        \begin{align}\label{eq:LF_profile}
            \mathcal{L}({\bf s}, J)
            &= \underset{{\bf c}:\|{\bf c}\|_0=J}{\max} ~ \frac{1}{\pi^{N}(\sigma_z^2 + \sigma_w^2)^{\|{\bf c}\|_0} (\sigma_w^2)^{N-\|{\bf c}\|_0}} \times \exp\left(- \sum_{i=1}^N \frac{|y_i - h_i {\bf u}_i^{\sf H}{\bf s}|^2}{c_i\sigma_z^2 + \sigma_w^2} \right) \nonumber \\
            &= \frac{1}{\pi^{N}(\sigma_z^2 + \sigma_w^2)^{J} (\sigma_w^2)^{N-J}} \times \exp\bigg(- \sum_{i=1}^{J} \frac{|{\sf sort}_i ({\bf e}({\bf s}))|^2}{\sigma_z^2 + \sigma_w^2} - \sum_{i=J+1}^N \frac{|{\sf sort}_i ({\bf e}({\bf s}))|^2}{\sigma_w^2} \bigg).
        \end{align}
        \hrulefill
        \vspace{-3mm}
    \end{figure*}
    Based on this analytical reformulation, the joint optimization problem in \eqref{eq:MLD} is transformed into a \textit{two-stage optimization problem.}
    Specifically, in the first stage, we determine the optimal number of jammed subcarriers, denoted as $\hat{J}({\bf s})$, for each candidate symbol vector ${\bf s}$ by maximizing the profile likelihood function defined in \eqref{eq:LF_profile}:
    \begin{align}\label{eq:MLD_lowcomp_J}
        \hat{J}({\bf s})
        &= \underset{J \in \{0, \dots, N\}}{\arg\!\max}~ \mathcal{L}({\bf s}, J) \nonumber \\
        &= \underset{J \in \{0, \dots, N\}}{\arg\!\max}~
        \frac{1}{(\sigma_z^2/\sigma_w^2 + 1)^{J}} \nonumber \\
        &~~\times\exp\bigg(- \sum_{i=1}^J \frac{|{\sf sort}_i ({\bf e}({\bf s}))|^2}{\sigma_z^2 + \sigma_w^2} - \sum_{i=J+1}^N \frac{|{\sf sort}_i ({\bf e}({\bf s}))|^2}{\sigma_w^2} \bigg).
    \end{align}
    In the second stage, after storing the maximum profile likelihood values $\mathcal{L}({\bf s}, \hat{J}({\bf s}))$ for all possible symbol vectors, the estimated symbol vector is obtained by solving the problem:
    \begin{align}\label{eq:MLD_lowcomp_s}
         \hat{\bf s}
        = \underset{{\bf s}\in\mathcal{X}_M^S}{\arg\!\max}~ \mathcal{L}({\bf s}, \hat{J}({\bf s})).
    \end{align}
    Additionally, the estimated number of jammed subcarriers is given by $\hat{J} = \hat{J}(\hat{\bf s})$.
    The overall procedure of the efficient MLD method is summarized in {\bf Algorithm~\ref{alg:lowcomp_mld}}.

    \begin{algorithm}[t]
        \caption{Efficient Maximum Likelihood Detection}
        \label{alg:lowcomp_mld}
        {\small
        {\begin{algorithmic}[1]
            \REQUIRE Received vector ${\bf y}$, channel matrix ${\bf H}$, spreading matrix ${\bf U}$,
            \STATEx \hspace{1.2em} modulation order $M$, jamming variance $\sigma_z^2$, and noise
            \STATEx \hspace{1.2em} variance $\sigma_w^2$
            \ENSURE Estimated symbol vector $\hat{\bf s}$ and estimated number of 
            \STATEx \hspace{2.3em}jammed subcarriers $\hat{J}$
            \FOR {each candidate symbol vector ${\bf s} \in \mathcal{X}_M^S$}
                \STATE Calculate the residual error vector ${\bf e}({\bf s}) = {\bf y} - {\bf H}{\bf U}{\bf s}$.
                \STATE Sort the elements of ${\bf e}({\bf s})$ in descending order of magnitude.
                \STATE Determine the optimal number of jammed subcarriers $\hat{J}({\bf s})$ for
                \STATEx \hspace{0.7em} the current ${\bf s}$ by solving \eqref{eq:MLD_lowcomp_J}.
                \STATE Store the likelihood value $L_{\max}({\bf s}) = \mathcal{L}({\bf s}, \hat{J}({\bf s}))$.
            \ENDFOR
            \STATE Determine the estimated symbol vector $\hat{\bf s}$ by solving \eqref{eq:MLD_lowcomp_s}.
            \STATE Set the estimated number of jammed subcarriers $\hat{J} = \hat{J}(\hat{\bf s})$.
        \end{algorithmic}}}
    \end{algorithm}
    
    We now analyze the computational complexity of the efficient MLD method.
    As outlined in {\bf Algorithm~\ref{alg:lowcomp_mld}}, the efficient MLD method iterates over all candidate symbol vectors ${\bf s} \in \mathcal{X}_M^S$.
    For each candidate ${\bf s}$, calculating the residual error vector requires a matrix-vector multiplication with a complexity of $\mathcal{O}(NS)$.
    Subsequently, sorting the magnitudes of the entries of the residual error vector requires a complexity of $\mathcal{O}(N \log N)$.
    Consequently, the overall computational complexity is dominated by $\mathcal{O}(M^S (NS + N \log N))$.
    Since the symbol vector length $S$ is at most $N$ (i.e., $S \le N$), the complexity is upper-bounded by $\mathcal{O}(M^S N^2)$.
    
    \textbf{Remark 3 (Mathematical Equivalence to Exhaustive MLD):}
    It is important to emphasize that the efficient MLD method achieves \textit{identical} detection performance to the exhaustive MLD method.
    This is because the joint optimization problem in \eqref{eq:MLD} is rigorously decomposed into the two-stage optimization problem involving \eqref{eq:MLD_lowcomp_J} and \eqref{eq:MLD_lowcomp_s} via the rearrangement inequality.
    Specifically, maximizing the profile likelihood function $\mathcal{L}({\bf s}, J)$ in \eqref{eq:LF_profile} is mathematically equivalent to maximizing the likelihood function $p({\bf y}|{\bf s},{\bf c})$ in \eqref{eq:LF_full} by implicitly optimizing the jamming indicator vector ${\bf c}$.
    Consequently, the efficient MLD method guarantees the global optimum of the joint optimization problem while reducing the computational complexity from $\mathcal{O}(M^S 2^N N)$ to $\mathcal{O}(M^S N^2)$.
    This substantial reduction in the computational burden improves the practical feasibility of the proposed anti-jamming modulation method as $N$ increases, since the complexity grows only quadratically with $N$ for a fixed number of transmitted bits $p$.

    \section{Analysis and Optimization of the Proposed Anti-Jamming Modulation}\label{Sec:Analysis}

    In this section, we analyze the detection performance of the proposed anti-jamming modulation method in the presence of the jamming signal.
    Based on this analysis, we derive the optimal modulation order to minimize detection errors according to jamming parameters.

    \subsection{BER Analysis}\label{Sec:BER}
    We characterize the theoretical BER of the proposed anti-jamming modulation method for an individual block under jamming attacks.
    To highlight the potential of our anti-jamming modulation method, we consider an idealized receiver employing the MLD method with complete knowledge of the channel matrix ${\bf H}$, the jamming indicator vector ${\bf c}$, the jamming variance $\sigma_z^2$, and the noise variance $\sigma_w^2$, which is referred to as the {\em genie-aided MLD}.
    Note that this complete knowledge is assumed solely for the analysis, whereas the MLD methods in Sec.~\ref{Sec:AJM} jointly estimate ${\bf c}$ along with the symbol vector ${\bf s}$, and the approximate MLD in Sec.~\ref{Sec:ApproxMLD} further estimates $\sigma_z^2$.
    The genie-aided MLD serves as a performance benchmark for these practical MLD methods, and the corresponding BER upper bound is referred to as the {\em genie-aided upper bound}.
    
    The BER for a given channel matrix ${\bf H}$ is expressed as
    \begin{align}\label{eq:BER0}
        P_e ({\bf H}) &= \frac{1}{p} \sum_{{\bf s} \in \mathcal{X}_M^S} \mathbb{P}( \hat{\bf s} \neq {\bf s}, {\bf s}~\text{is transmitted} \mid {\bf H}) d({\bf s},\hat{\bf s}),
    \end{align}
    where $d({\bf s},\hat{\bf s})$ represents the Hamming distance between the bit sequences associated with ${\bf s}$ and $\hat{\bf s}$.
    The BER in \eqref{eq:BER0} is upper bounded as follows:
    \begin{align}\label{eq:BER_upper}
        P_e ({\bf H}) \leq \frac{1}{p2^p}\sum_{{\bf s} \in \mathcal{X}_M^S}\sum_{\hat{\bf s} \neq {\bf s}}\mathbb{P}({\bf s} \rightarrow \hat{\bf s} \mid {\bf H}) d({\bf s},\hat{\bf s}),
    \end{align}
    where $\mathbb{P}({\bf s} \rightarrow \hat{\bf s} | {\bf H})$ represents the pairwise error probability (PEP), i.e., the probability that ${\bf s}$ is transmitted but detected as $\hat{\bf s}$.
    When employing the proposed anti-jamming modulation method at the transmitter and the genie-aided MLD at the receiver, the PEP is computed as \cite{Textbook}
    \begin{align}\label{eq:SER_Pair}
        &\mathbb{P}({\bf s} \rightarrow \hat{\bf s} \mid {\bf H}) \nonumber \\
        &= Q\left(\sqrt{\frac{({\bf U}{\bf s} - {\bf U}\hat{\bf s})^{\sf H}{\bf H}^{\sf H}{\bf \Sigma}^{-1}{\bf H}({\bf U}{\bf s} - {\bf U}\hat{\bf s})}{2}}\right),
    \end{align}
    where ${\bf \Sigma}$ is the jamming-plus-noise covariance matrix:
    \begin{align}
        {\bf \Sigma} = \sigma_z^2 {\sf diag}({\bf c})+ \sigma_w^2{\bf I}_N.
    \end{align}
    Note that unlike the PEP of the conventional OFDM scheme, the PEP in \eqref{eq:SER_Pair} additionally accounts for the effects of the spreading matrix ${\bf U}$ and the jamming indicator vector ${\bf c}$.
    The PEP can be simplified by analyzing the contributions from the jamming and noise components separately:
    \begin{align}\label{eq:SER_Pair2}
        \mathbb{P}({\bf s} \rightarrow \hat{\bf s} \mid {\bf H})
        &= Q\left(\sqrt{\frac{1}{2}\sum_{i=1}^N\frac{|h_i{\bf u}_{i}^{\sf H} ({\bf s} - \hat{\bf s})|^2}{c_i\sigma_z^2 + \sigma_w^2}}\right).
    \end{align}
    Our analysis in \eqref{eq:SER_Pair2} shows that as the jamming power approaches infinity (i.e., $\sigma_z^2 \to \infty$), the contribution of the terms associated with $c_i=1$ becomes negligible in the summation inside the Q-function.
    In this case, the PEP in \eqref{eq:SER_Pair2} is approximated as
    \begin{align}\label{eq:SER_Pair3}
        \mathbb{P}({\bf s} \rightarrow \hat{\bf s} \mid {\bf H})
        &\approx Q\left(\sqrt{\frac{1}{2}\sum_{i:c_i=0}\frac{|h_i{\bf u}_{i}^{\sf H} ({\bf s} - \hat{\bf s})|^2}{ \sigma_w^2}}\right).
    \end{align}
    This result implies that, even if the jamming power is infinite, the symbol vector transmitted via our anti-jamming modulation method can be detected without an error floor, provided that the number of unjammed subcarriers is sufficient to distinguish the symbol vectors (i.e., $N-\|{\bf c}\|_0 \ge S$).
    This observation is consistent with the solvability condition discussed in Sec.~\ref{Sec:Modulation}.
    
    Now, to characterize the average BER performance, we compute the average PEP, denoted as $\mathbb{P}({\bf s} \rightarrow \hat{\bf s})$, by averaging the conditional PEP in \eqref{eq:SER_Pair} over the distribution of the channel matrix ${\bf H}$.
    The result is given by
    \begin{align} \label{eq:UPEP}
        \mathbb{P}({\bf s} \rightarrow \hat{\bf s})
        =& \mathbb{E}_{\bf H}\big[ \mathbb{P}({\bf s} \rightarrow \hat{\bf s} \mid {\bf H}) \big] \nonumber \\
        \overset{(a)}{\approx} &\frac{1}{12}\left({\det\left({\bf I}_N + \frac{1}{4}{\bf K} {\bf \Sigma}^{-1} {\bf A}_{{\bf s},\hat{\bf s}}\right)}\right)^{-1} \nonumber \\
        &+ \frac{1}{4}\left({\det\left({\bf I}_N + \frac{1}{3}{\bf K} {\bf \Sigma}^{-1} {\bf A}_{{\bf s},\hat{\bf s}}\right)}\right)^{-1},
    \end{align}
    where $(a)$ follows from the Q-function approximation $Q(x) \approx \frac{1}{12} e^{-x^2/2} + \frac{1}{4} e^{-2x^2/3}$ \cite{Qfunction} when $h_i \sim \mathcal{CN}(0, \Omega_i)$.
    In \eqref{eq:UPEP}, ${\bf A}_{{\bf s},\hat{\bf s}}$ is a diagonal matrix defined as
    \begin{align}
        {\bf A}_{{\bf s},\hat{\bf s}} &= {\sf diag}({\bf U}{\bf s} - {\bf U}\hat{\bf s})^{\sf H}{\sf diag}({\bf U}{\bf s} - {\bf U}\hat{\bf s}) \nonumber \\
        &={\sf diag}(|{\bf u}_{1}^{\sf H} ({\bf s} - \hat{\bf s})|^2,\ldots,|{\bf u}_{N}^{\sf H} ({\bf s} - \hat{\bf s})|^2 ),
    \end{align}
    and ${\bf K} = \mathbb{E}[{\bf H}^{\sf H}{\bf H}] = {\sf diag}(\Omega_1,\ldots,\Omega_N)$ is the covariance matrix of ${\bf H}$.
    Since the matrices inside the determinant in \eqref{eq:UPEP} are diagonal, the determinants are computed for $r \in \{3, 4\}$ as
    \begin{align}
        \det\left({\bf I}_N + \frac{1}{r}{\bf K} {\bf \Sigma}^{-1} {\bf A}_{{\bf s},\hat{\bf s}}\right)
        &= \prod_{i=1}^N \left(1+\frac{\Omega_i |{\bf u}_{i}^{\sf H} ({\bf s} - \hat{\bf s})|^2}{r(c_i\sigma_z^2 + \sigma_w^2)}\right). \nonumber
    \end{align}
    Applying these results to the expression in \eqref{eq:UPEP} yields
    \begin{align} \label{eq:UPEP2}
        \mathbb{P}({\bf s} \rightarrow \hat{\bf s}) 
        &\approx \frac{1/12}{ \prod_{i=1}^N \left(1+\frac{\Omega_i |{\bf u}_{i}^{\sf H} ({\bf s} - \hat{\bf s})|^2}{4(c_i\sigma_z^2 + \sigma_w^2)}\right)} \nonumber \\
        &\quad + \frac{1/4}{ \prod_{i=1}^N \left(1+\frac{\Omega_i |{\bf u}_{i}^{\sf H} ({\bf s} - \hat{\bf s})|^2}{3(c_i\sigma_z^2 + \sigma_w^2)}\right)}.
    \end{align}
    Consequently, the theoretical BER is approximately upper bounded by
    \begin{align}\label{eq:BER_upper2}
        P_e &= \mathbb{E}_{\bf H}\big[P_e({\bf H})\big] \nonumber \\
        &\leq \frac{1}{p2^p}\sum_{{\bf s} \in \mathcal{X}_M^S}\sum_{\hat{\bf s} \neq {\bf s}}\mathbb{P}({\bf s} \rightarrow \hat{\bf s} ) d({\bf s},\hat{\bf s}) \nonumber \\
        &\approx \frac{1}{p2^p}\sum_{{\bf s} \in \mathcal{X}_M^S}\sum_{\hat{\bf s} \neq {\bf s}} \Bigg\{\frac{d({\bf s},\hat{\bf s})/12}{ \prod_{i=1}^N \left(1+\frac{\Omega_i |{\bf u}_{i}^{\sf H} ({\bf s} - \hat{\bf s})|^2}{4(c_i\sigma_z^2 + \sigma_w^2)}\right)}\nonumber \\
        &\qquad\qquad\qquad\qquad + \frac{d({\bf s},\hat{\bf s})/4}{ \prod_{i=1}^N \left(1+\frac{\Omega_i|{\bf u}_{i}^{\sf H} ({\bf s} - \hat{\bf s})|^2}{3(c_i\sigma_z^2 + \sigma_w^2)}\right)}\Bigg\}\nonumber \\
        &\triangleq P_{e,upper} ({\bf c},\sigma_z^2,{\bm \Omega}),
    \end{align}
    where ${\bm \Omega} = [\Omega_1, \dots, \Omega_N]^{\sf T}$ is the channel power vector.
    As shown in \eqref{eq:BER_upper2}, the approximate upper bound of the theoretical BER is expressed as a function of the jamming indicator vector ${\bf c}$, the jamming variance $\sigma_z^2$, and the channel power vector ${\bm \Omega}$.
    This analysis reveals how the channel and jamming environments influence the performance of the proposed anti-jamming modulation method, providing a theoretical foundation for understanding its advantages and behavior.

    \subsection{Modulation Order Optimization}\label{Sec:Optimization}
    
    As indicated in \eqref{eq:BER_upper2}, the approximate upper bound $P_{e,upper} ({\bf c},\sigma_z^2,{\bm \Omega})$ is also determined by the choice of the constellation set ${\mathcal X}_M^S$.
    Since the channel and jamming environments are not controllable, we optimize the modulation order $M$, the only adjustable parameter, by using this approximate upper bound as the objective function.
    Note that as the proposed modulation method maintains a constant spectral efficiency regardless of $M$, minimizing the BER is strictly equivalent to maximizing the effective throughput.

    In the proposed modulation method, $p$ bits are first mapped to an $S$-dimensional symbol vector ${\bf s}$ and then transformed into an $N$-dimensional modulated vector ${\bf x}$ utilizing the spreading matrix ${\bf U} \in \mathbb{C}^{N\times S}$.
    Thus, the modulation order must satisfy $M = 2^{p/S}$ for $1\leq S\leq N$.
    From this constraint, a candidate set for the modulation order is constructed as ${\mathcal M} = \{M_1, M_2, \dots, M_L\}$, where $M_i = 2^{p/S_i}$ for $1\leq S_i \leq N$ such that $p/S_i$ is an integer.
    The optimization problem to determine the optimal modulation order $M^\star$ is then given by:
    \begin{align}\label{eq:M_opt}
        M^\star = \underset{M_i \in \mathcal{M}}{\arg\!\min}~P_{e,upper} ({\bf c},\sigma_z^2,{\bm \Omega}).
    \end{align}

    The optimization problem formulated in \eqref{eq:M_opt} provides the optimal modulation order for a specific jamming indicator vector ${\bf c}$ within an individual block.
    However, directly solving this problem in practical systems is undesirable.
    The jamming indicator vector ${\bf c}$ varies across multiple blocks due to both the jamming scenarios and the deinterleaving patterns, which implies that the optimal modulation order would need to change for each block.
    Consequently, applying the optimization for each block imposes a significant system burden.
    For instance, if the optimization is performed at the receiver, communicating the optimal modulation order for each block would introduce substantial feedback overhead.

    To address this challenge, we propose a practical optimization problem to determine the modulation order.
    The core idea is to determine the best modulation order in an average sense across multiple blocks according to the channel and jamming environments.
    We assume that the channel covariance matrix ${\bf K} = {\sf diag}(\Omega_1, \dots, \Omega_N)$ is approximately given by ${\bf K} \approx \bar{\Omega} {\bf I}_N$, where $\bar{\Omega}$ represents the average channel power across multiple blocks.
    We further assume that the average number of jammed subcarriers in each block is denoted as $J_{\rm avg}$.
    Under these assumptions, for any jamming indicator vector ${\bf c}$ with $\|{\bf c}\|_0=J_{\rm avg}$, the approximate upper bound in \eqref{eq:BER_upper2} can be further upper bounded using the rearrangement inequality as given in \eqref{eq:BER_upper_avg}.
    \begin{figure*}
    \begin{align} \label{eq:BER_upper_avg}
        P_{e,upper}({\bf c},\sigma_z^2,{\bm \Omega}) &\leq \frac{1}{p2^p}\sum_{{\bf s} \in \mathcal{X}_M^S}\sum_{\hat{\bf s} \neq {\bf s}} \Bigg\{\frac{d({\bf s},\hat{\bf s})/12}{ \prod_{i=1}^{J_{\rm avg}} \left(1+\frac{\bar{\Omega} |{\sf sort}_i({\bf U}({\bf s} - \hat{\bf s}))|^2}{4(\sigma_z^2 + \sigma_w^2)}\right) \prod_{i=J_{\rm avg}+1}^N \left(1+\frac{\bar{\Omega} |{\sf sort}_i({\bf U}({\bf s} - \hat{\bf s}))|^2}{4\sigma_w^2}\right)}\nonumber \\
        &\qquad\qquad\qquad\qquad + \frac{d({\bf s},\hat{\bf s})/4}{ \prod_{i=1}^{J_{\rm avg}} \left(1+\frac{\bar{\Omega} |{\sf sort}_i({\bf U}({\bf s} - \hat{\bf s}))|^2}{3(\sigma_z^2 + \sigma_w^2)}\right) \prod_{i=J_{\rm avg}+1}^N \left(1+\frac{\bar{\Omega} |{\sf sort}_i({\bf U}({\bf s} - \hat{\bf s}))|^2}{3\sigma_w^2}\right)}\Bigg\}\nonumber \\
        & \triangleq \bar{P}_{e,upper}(J_{\rm avg},\sigma_z^2,\bar{\Omega}).
    \end{align}
    \hrulefill
    \vspace{-3mm}
    \end{figure*}
    Utilizing this average upper bound $\bar{P}_{e,upper}(J_{\rm avg},\sigma_z^2,\bar{\Omega})$, the optimal modulation order is determined by:
    \begin{align}\label{eq:M_opt2}
        M^\star = \underset{M_i \in \mathcal{M}}{\arg\!\min}~ \bar{P}_{e,upper}(J_{\rm avg},\sigma_z^2,\bar{\Omega}).
    \end{align}
    The optimal modulation order determined from \eqref{eq:M_opt2} remains valid as long as the channel and jamming environments do not change significantly.
    Although our optimization requires knowledge of the jamming parameters (i.e., $J_{\rm avg}$ and $\sigma_z^2$), these parameters can be estimated before the modulation order optimization, which will be clarified in Sec.~\ref{Sec:Adaptive}.

    Our optimization adaptively selects the optimal modulation order according to the channel and jamming environments.
    As described in Sec.~\ref{Sec:Modulation}, selecting a higher modulation order $M$ (i.e., smaller $S$) leads to a higher spreading gain of $N/S$, but reduces the minimum Euclidean distance between constellation points, making the system more susceptible to noise.
    For instance, when the number of jammed subcarriers is large while satisfying $J < N$ with high jamming power, the proposed optimization in \eqref{eq:M_opt2} selects a high modulation order to increase the spreading gain.
    Conversely, when $J$ is small or the jamming power is low, it selects a low modulation order to increase the minimum Euclidean distance.
    This implies that our modulation order optimization successfully considers the effects of both the jamming signal and the noise based on the BER analysis in Sec.~\ref{Sec:BER}.

    \section{Proposed Jamming-Adaptive Communication Framework}\label{Sec:Adaptive}
    In practical systems, the jamming parameters, such as the average number of jammed subcarriers $J_{\rm avg}$ and the jamming variance $\sigma_z^2$, are unknown prior to transmission.
    This hinders the use of the proposed anti-jamming modulation and demodulation methods in Sec.~\ref{Sec:AJM} as well as the modulation order optimization in Sec.~\ref{Sec:Analysis}.
    To address this challenge, in this section, we propose a jamming-adaptive communication framework that enables our modulation and demodulation methods to adapt to dynamic jamming without prior knowledge of the jamming parameters.
    
    \subsection{Approximate MLD Method with Unknown Jamming Variance}\label{Sec:ApproxMLD}
    To establish the jamming-adaptive communication framework, we first develop an {\em approximate} MLD method for an individual block without prior knowledge of the jamming parameters.
    Our approach is based on the GLRT principle, where the unknown jamming variance is estimated for each hypothesis of the symbol vector and the number of jammed subcarriers.
    
    Recall from Sec.~\ref{Sec:LowcompMLD} that the residual error vector associated with a candidate symbol vector ${\bf s} \in \mathcal{X}_M^S$ is defined as ${\bf e}({\bf s}) = {\bf y} - {\bf H} {\bf U} {\bf s}$.
    Under the hypothesis that the candidate symbol vector ${\bf s}$ matches the true symbol vector ${\bf s}_{\rm true}$, the residual error vector consists only of the jamming and noise components.
    In this case, assuming that the number of jammed subcarriers is given by $J$, the expectation of the squared norm of the residual error vector is derived as
    \begin{align}\label{eq:expect_residual}
        \mathbb{E}[\|{\bf e}({\bf s}) \|^2] 
        &= \sum_{i=1}^N \mathbb{E}[|h_i {\bf u}_i^{\sf H}({\bf s}_{\rm true} - {\bf s} ) + c_iz_i + w_i|^2] \nonumber \\
        &= \sum_{i=1}^N c_i \mathbb{E}[ |z_i|^2] + \sum_{i=1}^N \mathbb{E}[ |w_i|^2] \nonumber \\
        &= J\sigma_z^2 + N\sigma_w^2.
    \end{align}
    Since the expected value is unavailable at the receiver, we estimate the jamming variance by replacing the expected value in \eqref{eq:expect_residual} with the instantaneous value $\|{\bf e}({\bf s}) \|^2$.
    This leads to the conditional estimate of the jamming variance for the hypothesis $\{{\bf s}, J\}$:
    \begin{align}\label{eq:var_approx}
        \hat{\sigma}_z^2({\bf s}, J) =
        \begin{cases}
         \max\left(\frac{\|{\bf e}({\bf s})\|^2 - N \sigma_w^2}{J}, 0\right), &~\text{if } J > 0, \\
         0, &~\text{if } J = 0.
        \end{cases}
    \end{align}
    Note that $\hat{\sigma}_z^2({\bf s}, J)$ is set to zero when $\|{\bf e}({\bf s})\|^2 \leq N \sigma_w^2$, which is consistent with the jamming-free scenario.
    
    We formulate the approximate MLD method by substituting the unknown jamming variance in the likelihood function with the conditional estimate in \eqref{eq:var_approx}.
    By incorporating the sorted structure of the residual errors as discussed in Sec.~\ref{Sec:LowcompMLD}, the approximate MLD method estimates the symbol vector and the number of jammed subcarriers by maximizing the \textit{approximate profile likelihood function} $\tilde{\mathcal{L}}({\bf s}, J)$:
    \begin{align}\label{eq:approx_opt}
         &(\hat{\bf s}, \hat{J}) \nonumber \\
         &= \underset{{\bf s}\in\mathcal{X}_M^S, J \in \{0, \dots, N\}}{\arg\!\max}~\tilde{\mathcal{L}}({\bf s}, J) \nonumber \\
         &= \underset{{\bf s}\in\mathcal{X}_M^S, J \in \{0, \dots, N\}}{\arg\!\max}~\frac{1}{(\hat{\sigma}_z^2({\bf s}, J)/\sigma_w^2 + 1)^{J}} \nonumber \\
         &\quad \times \exp\Bigg( - \sum_{i=1}^J \frac{|{\sf sort}_i ({\bf e}({\bf s}))|^2}{\hat{\sigma}_z^2({\bf s}, J) + \sigma_w^2} - \sum_{i=J+1}^N \frac{|{\sf sort}_i ({\bf e}({\bf s}))|^2}{\sigma_w^2} \Bigg).
    \end{align}
    Since the optimization problem in \eqref{eq:approx_opt} shares the same structural properties as the efficient MLD method, it can be solved efficiently using the same two-stage procedure described in {\bf Algorithm~\ref{alg:lowcomp_mld}}.
    This ensures that the approximate MLD method maintains the computational complexity of $\mathcal{O}(M^S N^2)$.
    
    Although the estimator in \eqref{eq:var_approx} assumes that ${\bf s}$ is the true symbol vector, this logic effectively penalizes incorrect hypotheses.
    Specifically, if ${\bf s} \neq {\bf s}_{\rm true}$, the residual error vector contains the non-zero signal mismatch term ${\bf H}{\bf U}({\bf s}_{\rm true} - {\bf s})$.
    This implies that the residual error on the $i$-th subcarrier includes the component $|h_i {\bf u}_i^{\sf H}({\bf s}_{\rm true} - {\bf s} )|$, resulting in a relatively large value of $\|{\bf e}({\bf s})\|^2$.
    Consequently, the jamming variance $\hat{\sigma}_z^2({\bf s}, J)$ is overestimated, which in turn reduces the likelihood value in \eqref{eq:approx_opt}.
    This mechanism ensures that the correct symbol vector is selected with high probability even without prior knowledge of the jamming variance.

    It is worth noting that although the proposed efficient and approximate MLD methods are derived under the i.i.d. complex Gaussian assumption, they identify the heavily jammed subcarriers through the sorted residual errors and assign a lower weight to the corresponding observations.
    Since this mechanism relies on the magnitude of the residual errors rather than the specific distribution of the jamming signal, these methods remain effective even when the jamming signal deviates from this assumption, as evaluated in Fig.~\ref{Fig_7}(b).

    \subsection{Jamming-Adaptive Communication Framework}
    
    \begin{figure}[t]
        \centering
        \includegraphics[width=0.98\columnwidth]{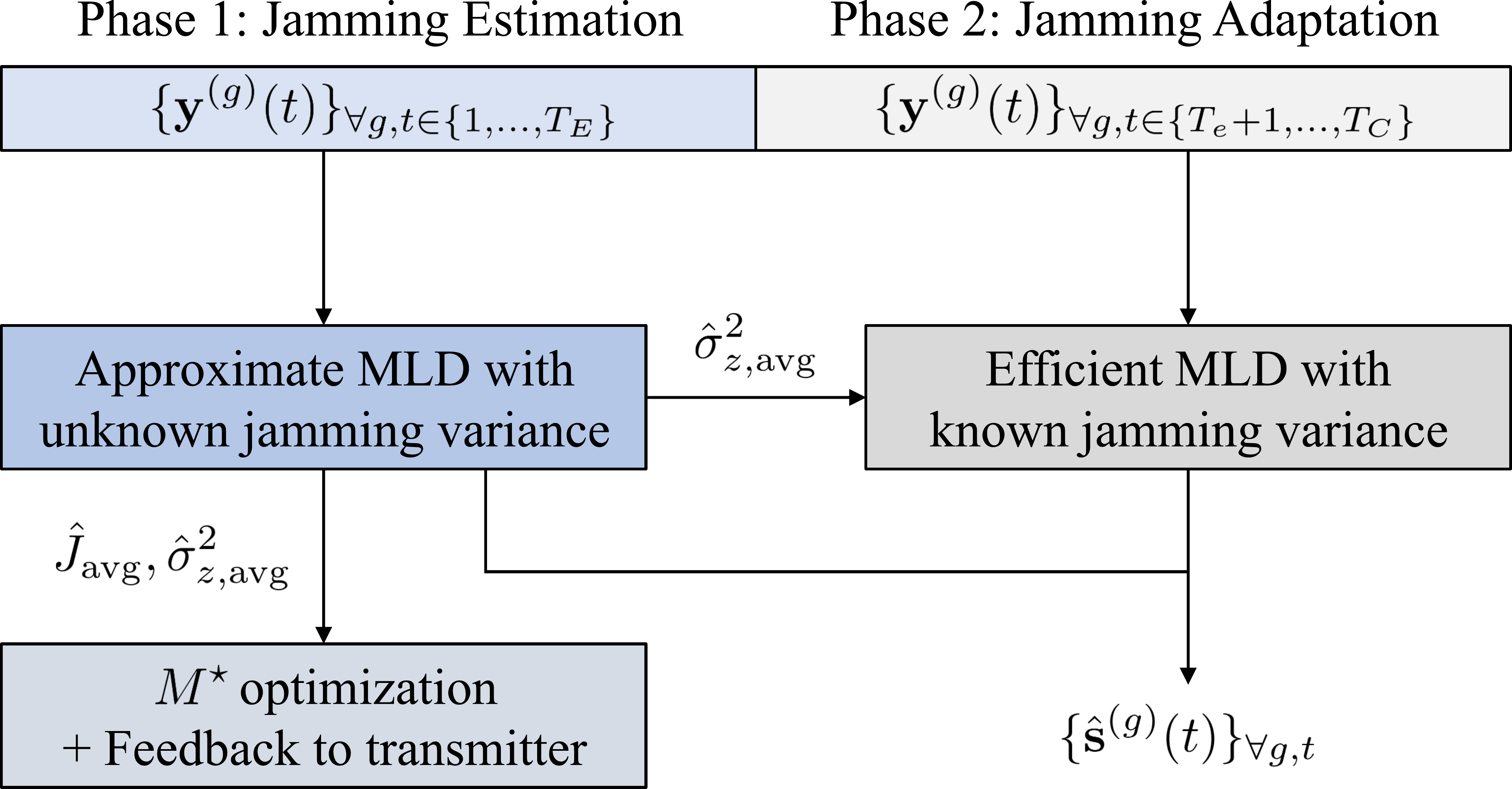}
        \vspace{-2mm}
        \caption{Illustration of the jamming-adaptive communication framework.}
        \vspace{-3mm}
        \label{Fig_3}
    \end{figure}
    
    Extending the discussion from the operation for an individual block in Sec.~\ref{Sec:ApproxMLD}, this subsection presents the overall jamming-adaptive communication framework that operates over multiple blocks within a {\em transmission cycle} of duration $T_C$ (i.e., $T_C$ consecutive OFDM symbols).
    The overall structure and procedure of this framework are illustrated in Fig.~\ref{Fig_3}.
    In practical systems, we consider dynamic jamming scenarios where the jamming scenario or the jamming parameters vary over time.
    To address such dynamic jamming scenarios, the transmission cycle duration $T_C$ is designed as an adjustable parameter.
    By adjusting $T_C$ according to the dynamic jamming scenarios, our framework ensures that the average\footnote{We only assume that the {\em average} number of jammed subcarriers across the blocks remains unchanged, while allowing that the number of jammed subcarriers can vary over different blocks depending on the jamming scenarios and the deinterleaving patterns.} number of jammed subcarriers and the jamming variance remain effectively constant within a transmission cycle.
    Each transmission cycle consists of two phases: (i) a {\em jamming estimation} phase, which occupies the first $T_E$ OFDM symbols, and (ii) a {\em jamming adaptation} phase, which spans the remaining $T_C - T_E$ OFDM symbols.
    
    During the jamming estimation phase, the transmitter utilizes the modulation order $M$ determined in the previous transmission cycle.
    Note that for the initial transmission cycle, $M$ is set to 4 (i.e., 4-QAM) to ensure robust performance under unknown jamming attacks.
    Since the jamming parameters may have changed from the previous cycle, the receiver estimates both the symbol vector and the number of jammed subcarriers for each block employing the approximate MLD method described in Sec.~\ref{Sec:ApproxMLD}.
    Let $\hat{\bf s}^{g}(t)$ and $\hat{J}^{g}(t)$ denote the estimated symbol vector and the estimated number of jammed subcarriers, respectively, for the received signal ${\bf y}^g(t)$ associated with block $g$ of the $t$-th OFDM symbol.
    
    At the end of the jamming estimation phase, the receiver estimates the average number of jammed subcarriers, $\hat{J}_{\rm avg}$, and the jamming variance, $\hat{\sigma}_{z,{\rm avg}}^2$, based on the estimation results $\{\hat{\bf s}^{g}(t), \hat{J}^{g}(t)\}_{{\forall g},t\in\{1,\dots,T_E\}}$ obtained from all blocks.
    The average number of jammed subcarriers is estimated as the most frequently occurring value among the estimated numbers of jammed subcarriers gathered during this phase:
    \begin{align}\label{eq:J_avg}
        \hat{J}_{\rm avg} = \underset{0 \leq J \leq N}{\arg\!\max} \left( \sum_{t=1}^{T_E} \sum_{g=1}^{G} 1[\hat{J}^{g}(t) = J] \right),
    \end{align}
    where $1[\cdot]$ is the indicator function.
    The jamming variance is estimated as the average power of the residual errors, excluding the noise power contribution from the jammed subcarriers.
    This estimation is performed over multiple blocks and OFDM symbols as follows:
    \begin{align}\label{eq:sigma_z}
        \hat{\sigma}_{z,{\rm avg}}^2 &= \frac{\sum_{t=1}^{T_E}\sum_{g=1}^G \max\left(\Delta^g(t), 0\right)}{\sum_{t=1}^{T_E}\sum_{g=1}^G \hat{J}^g(t)},
    \end{align}
    where $\Delta^g(t)$ represents the instantaneous jamming power estimate for each block, defined as:
    \begin{align}\label{eq:delta}
        \Delta^g(t) &=
        \begin{cases}
        \left\|{\bf y}^g(t) - {\bf H}^g(t){\bf U}\hat{\bf s}^g(t)\right\|^2 - N\sigma_w^2, & \text{if } \hat{J}^g(t) > 0, \\
        0, & \text{if } \hat{J}^g(t) = 0.
        \end{cases}
    \end{align}
    
    Using the estimated jamming parameters, the receiver determines the new optimal modulation order $M^\star$ by solving the modulation order optimization problem in \eqref{eq:M_opt2}.
    The receiver then feeds back $M^\star$ to the transmitter.
    Note that this feedback requires only $\lceil \log_2 |{\mathcal M}| \rceil$ bits once per transmission cycle, and thus the feedback overhead is negligible.
    
    During the jamming adaptation phase, the transmitter utilizes the updated modulation order $M^\star$ for data transmission.
    Correspondingly, the receiver estimates the symbol vectors for each block employing the efficient MLD method in Sec.~\ref{Sec:LowcompMLD} with the estimated jamming variance $\hat{\sigma}_{z,{\rm avg}}^2$.
    This updated modulation order $M^\star$ is maintained and used as the initial modulation order for the subsequent transmission cycle, establishing a recursive update structure.
    The overall procedure of the proposed jamming-adaptive communication framework for a single transmission cycle is summarized in {\bf Algorithm~\ref{alg:adaptive_framework}}.
    
    \begin{algorithm}[t]
        \caption{Jamming-Adaptive Communication Framework}
        \label{alg:adaptive_framework}
        {\small
        \begin{algorithmic}[1]
            \REQUIRE Received vectors $\{{\bf y}^{g}(t)\}_{\forall g,t}$, channel matrices $\{{\bf H}^{g}(t)\}_{\forall g,t}$,
            \Statex \hspace{1.2em} unitary matrix ${\bf U}_0$, current modulation order $M$, and noise 
            \Statex \hspace{1.2em} variance $\sigma_w^2$
            \ENSURE Estimated symbol vectors $\{\hat{\bf s}^{g}(t)\}_{\forall g,t}$ and updated modu-
            \Statex \hspace{2.0em} lation order $M^\star$
    
            \vspace{0.5em}
            \STATE \textbf{Phase 1: Jamming Estimation}
            \FOR {$t = 1$ to $T_E$}
                \FOR {$g = 1$ to $G$}
                    \STATE Estimate $(\hat{\bf s}^{g}(t),\hat{J}^{g}(t))$ employing the approximate MLD in 
                    \Statex \hspace{1.6em} \eqref{eq:approx_opt}.
                \ENDFOR
            \ENDFOR
            \STATE Compute $\hat{J}_{\rm avg}$ and $\hat{\sigma}_{z,{\rm avg}}^2$ using \eqref{eq:J_avg} and \eqref{eq:sigma_z}.
            \STATE Determine the optimal modulation order $M^\star$ from \eqref{eq:M_opt2}.
            \STATE Feed back the updated modulation order $M^\star$ to the transmitter.
    
            \vspace{0.5em}
            \STATE \textbf{Phase 2: Jamming Adaptation}
            \FOR{$t = T_E + 1$ to $T_C$}
                \FOR {$g = 1$ to $G$}
                    \STATE Estimate $\hat{\bf s}^{g}(t)$ employing the efficient MLD (Algorithm~\ref{alg:lowcomp_mld})
                    \Statex \hspace{1.7em} with estimated jamming variance $\hat{\sigma}_{z,{\rm avg}}^2$.
                \ENDFOR
            \ENDFOR
            \STATE \textbf{Return} $\{\hat{\bf s}^{g}(t)\}_{\forall g,t}$ and $M^\star$.
        \end{algorithmic}}
    \end{algorithm}

    \section{Simulation Results}\label{Sec:Simulation}

    In this section, we demonstrate the superiority of the proposed anti-jamming OFDM scheme compared to existing OFDM schemes by evaluating the BER and effective throughput.
    To verify the robustness of our scheme, we conduct Monte Carlo simulations under various jamming scenarios.

    \subsection{Simulation Setup}\label{Sec:Setup}

    \begin{table}[t]
        \renewcommand{\arraystretch}{1.3}
        \caption{Simulation Parameters}
        \label{Tab:Params}
        \centering
        \begin{tabular}{ll}
        \toprule
        \textbf{Parameter} & \textbf{Value} \\
        \midrule
        Number of subcarriers, $K$ & $512$ \\
        Number of OFDM symbols, $T$ & $200$ \\
        Channel model & Rayleigh fading, i.i.d. $\mathcal{CN}(0,1)$ \\
        Additive noise & AWGN \\
        SNR, $1/\sigma_w^2$ & $\{0, 5, \ldots, 20\}$~dB \\
        SJR, $1/\sigma_z^2$ & $\{-40, -30, \ldots, 20\}$~dB \\
        Jamming ratio, $\rho$ & $(0,\, 1]$ \\
        Transmission cycle, $T_C$ & $\{28, 56, 112\}$ \\
        Estimation phase, $T_E$ & $T_C/2$ \\
        \bottomrule
        \end{tabular}
        \vspace{-3mm}
    \end{table}

    In our simulations, we consider an OFDM system with $K = 512$ subcarriers and $T = 200$ OFDM symbols.
    The subcarriers are divided into blocks of size $N$, resulting in a total of $G = \lceil K / N \rceil$ blocks.
    If $K$ is not divisible by $N$, zero-padding is applied to the last block to ensure uniform block sizes.
    The channel is modeled as Rayleigh fading, where the CFR follows an independent and identically distributed (i.i.d.) $\mathcal{CN}(0, 1)$ distribution, and the additive noise is modeled as AWGN.
    The signal-to-noise ratio (SNR) is defined as $\text{SNR} = 1/\sigma_w^2$.
    Unless otherwise stated, we assume that the receiver has perfect knowledge of the CFR and the noise variance, and frequency interleaving is employed as explained in Sec.~\ref{Sec:System}.

    Various jamming scenarios are considered in the simulations.
    The signal-to-jamming ratio (SJR) is defined as $\text{SJR} = 1/\sigma_z^2$.
    The jamming ratio $\rho \in (0, 1]$ denotes the fraction of time or frequency resources corrupted by the jamming signal.
    Unless otherwise stated, the jamming signal follows the Gaussian model described in Sec.~\ref{Sec:System}.
    The details of each jamming scenario are described as follows:
    \begin{itemize}
        \item \textbf{Partial-band Jamming:}
        The jammer attacks a specific fraction $\rho$ of contiguous subcarriers, while the others remain jamming-free.
        
        \item \textbf{Pulse Jamming:}
        The jammer attacks all subcarriers simultaneously with a duty cycle $\rho$, where it is active for a fraction $\rho$ of the OFDM symbols.
        
        \item \textbf{Random Jamming:}
        The jammer randomly selects a subset of subcarriers with a probability of $\rho$ to attack in each OFDM symbol.
        
        \item \textbf{Barrage Jamming:}
        The jammer continuously attacks all subcarriers, which corresponds to the case where $\rho = 1$.
    \end{itemize}
    Note that the deinterleaving process disperses burst errors into sparse errors regardless of the jamming scenario.
    The simulation parameters are summarized in Table~\ref{Tab:Params}.

    For performance comparison, we consider the following OFDM schemes.
    To ensure a fair comparison, all schemes transmit the same $m$ information bits across $K$ subcarriers under the same total transmit power, thereby achieving the same spectral efficiency.
    Unless otherwise stated, we set $m=512$, resulting in a spectral efficiency of $1$~bps/Hz, while $m=256$ is additionally considered in Fig.~\ref{Fig_8} to evaluate a spectral efficiency of $0.5$~bps/Hz.
    To verify their practicality, the computational complexity per block for the demodulation process is also summarized in Table~\ref{Tab:Scheme}.

    \begin{itemize}
        \item \textbf{AJ-OFDM:}
        The proposed anti-jamming OFDM scheme employing the modulation method in Sec.~\ref{Sec:AJM}.
        For each block, $p$ bits are transformed into an $N$-dimensional modulated vector utilizing the modulation order $M$ and the spreading matrix $\mathbf{U}$.
        Unless otherwise stated, the efficient MLD in Sec.~\ref{Sec:LowcompMLD} is employed assuming knowledge of the jamming variance $\sigma_z^2$.

        \item \textbf{AJ-OFDM-Adapt:}
        The proposed jamming-adaptive communication framework described in Sec.~\ref{Sec:Adaptive}.
        It operates over a transmission cycle of $T_C$ OFDM symbols, where the jamming estimation phase is set to $T_E = T_C/2$.
        During the jamming estimation phase, it employs the approximate MLD and estimates the jamming parameters.
        Subsequently, the optimal modulation order $M^\star$ is determined in \eqref{eq:M_opt2} for the transmission.
        Then, during the jamming adaptation phase, the efficient MLD is employed with the estimated jamming variance.

        \item \textbf{Conventional OFDM:}
        The standard OFDM scheme, where for each block, $p=1$ bit is transmitted across $N=1$ subcarrier using BPSK ($M=2$) to achieve the target spectral efficiency.

        \item \textbf{OFDM-IM:}
        The OFDM scheme with index modulation, where for each block, $p=4$ bits are transmitted across $N=4$ subcarriers by activating $N_A=2$ subcarriers using BPSK to achieve the target spectral efficiency.
        This scheme has the potential to mitigate jamming attacks as it selectively activates subcarriers based on index bits, reducing the likelihood that active subcarriers are corrupted by the jamming signal.

        \item \textbf{FH-OFDM:}
        The frequency hopping OFDM scheme utilizing a hopping pattern known to both the transmitter and receiver.
        For a single block, $p=512$ bits are transmitted across $N=512$ subcarriers by activating $N_A=256$ subcarriers using 4-QAM ($M=4$) to achieve the target spectral efficiency.
        Similar to OFDM-IM, this scheme mitigates jamming attacks by selectively activating subcarriers based on the hopping pattern.

        \item \textbf{WHT-OFDM:}
        The OFDM scheme utilizing a truncated WHT matrix for frequency spreading.
        For a single block, $p=512$ bits are mapped to 256 symbols using 4-QAM and transformed into an $N=512$-dimensional modulated vector utilizing a $512 \times 256$ WHT matrix to achieve the target spectral efficiency.
        This scheme exploits a spreading gain of 2 to enhance robustness against jamming attacks while employing an MMSE detector.
    \end{itemize}

    \begin{table}[t]
        \renewcommand{\arraystretch}{1.3}
        \caption{Computational Complexity Comparison per Block}
        \label{Tab:Scheme}
        \centering
        \begin{threeparttable}
        \begin{tabular}{llc}
        \toprule
        \textbf{Scheme} & \textbf{Detector} & \textbf{Complexity} \\
        \midrule
        \multirow{2}{*}{\textbf{AJ-OFDM}$^{\dagger}$} & Exhaustive MLD & $\mathcal{O}(M^S 2^N N)$ \\
        & Efficient / Approximate MLD & $\bm{\mathcal{O}(M^S N^2)}$ \\
        \midrule
        Conv. OFDM & MLD & $\mathcal{O}(MN)$ \\
        OFDM-IM & MLD & $\mathcal{O}(\binom{N}{N_A} M^{N_A} N)$ \\
        FH-OFDM & MLD & $\mathcal{O}(MN_A)$ \\
        WHT-OFDM & MMSE & $\mathcal{O}(N^2)$ \\
        \bottomrule
        \end{tabular}
        \begin{tablenotes}
            \item[$\dagger$] \footnotesize{Note that $M^S = 2^p$ is a constant when $p$ is fixed.}
        \end{tablenotes}
        \end{threeparttable}
    \end{table}

    \subsection{Performance Evaluation}

    \begin{figure}[t!]
        \centering
        \subfloat[$(p, N, M) = (4, 4, 4)$]{\includegraphics[width=0.49\columnwidth]{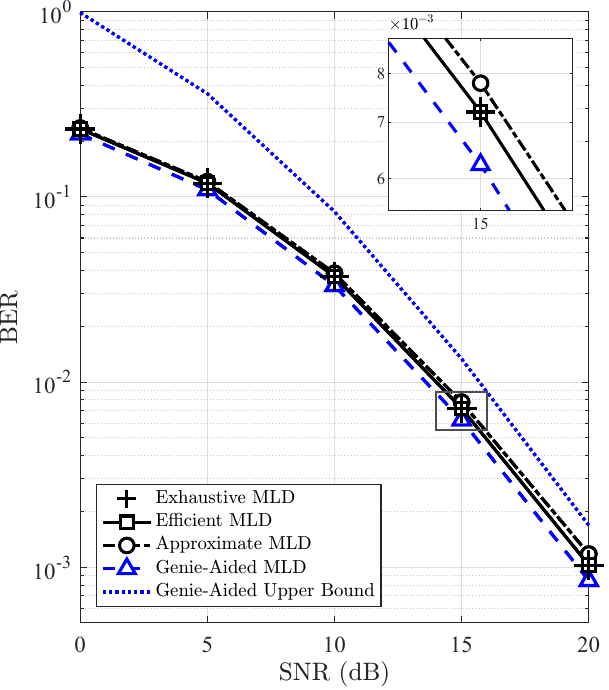}%
        \label{fig:Fig4_a}}
        \hfill
        \subfloat[$(p, N, M) = (6, 6, 8)$]{\includegraphics[width=0.49\columnwidth]{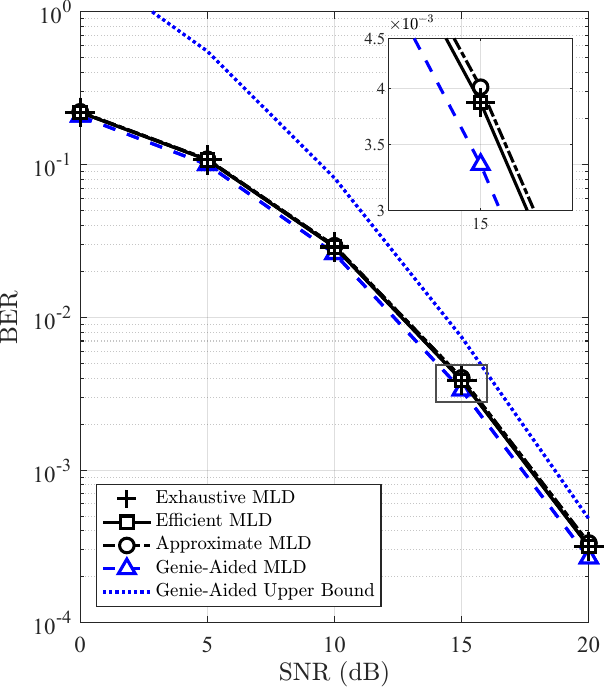}%
        \label{fig:Fig4_b}}
        \vspace{-1mm}
        \caption{BER comparison of AJ-OFDM with the exhaustive, efficient, approximate, and genie-aided MLDs and the genie-aided upper bound versus SNR under partial-band jamming ($\rho = 0.5$, $\text{SJR} = -20$~dB).}
        \label{Fig_4}
        \vspace{-3mm}
    \end{figure}

\begin{figure}[t!]
        \vspace{-3mm}
        \centering
        \subfloat[Partial-band and pulse jamming with Gaussian jamming signals]{\includegraphics[width=0.49\columnwidth]{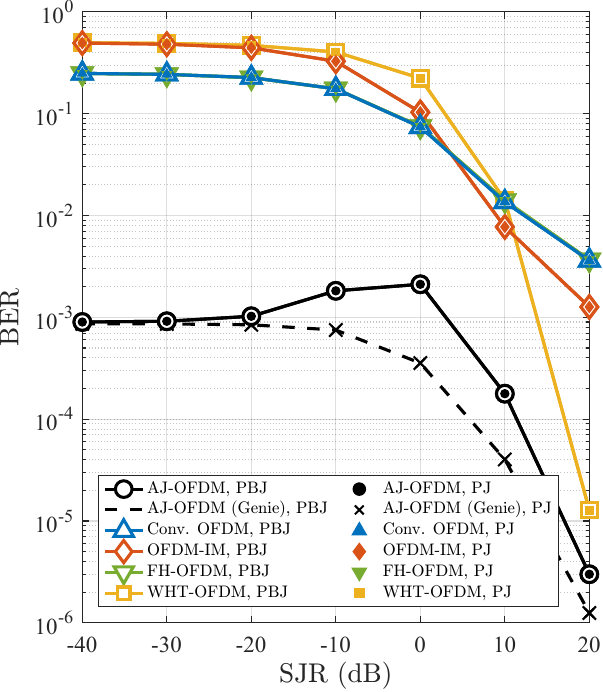}%
        \label{fig:joint}}
        \hfill
        \subfloat[Reactive jamming with Gaussian and impulsive jamming signals]{\includegraphics[width=0.49\columnwidth]{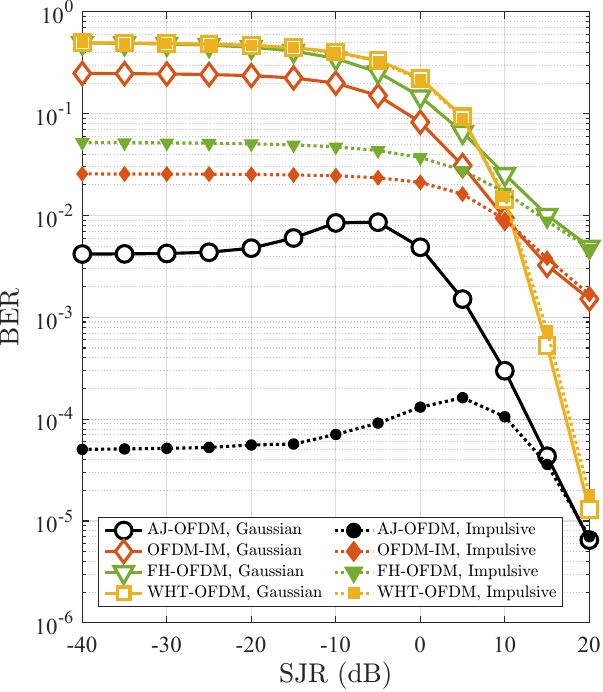}%
        \label{fig:waveform}}
        \vspace{-1mm}
        \caption{BER comparison of AJ-OFDM with $(p, N, M) = (4, 4, 4)$ and existing OFDM schemes versus SJR under various jamming scenarios and types ($\rho=0.5,~\text{SNR}=20$~dB), where joint time-frequency interleaving is employed in (a) and AJ-OFDM employs the approximate MLD in (b).}
        \label{Fig_7}
        \vspace{-3mm}
    \end{figure}

    In Fig.~\ref{Fig_4}, we compare the BERs of AJ-OFDM for different SNRs under partial-band jamming with $\rho = 0.5$ and $\text{SJR} = -20$~dB, where $(p, N, M) = (4, 4, 4)$ and $(6, 6, 8)$ are considered in Figs.~\ref{Fig_4}(a) and~\ref{Fig_4}(b), respectively.
    As shown in Fig.~\ref{Fig_4}, the BER curves of the efficient MLD perfectly match those of the exhaustive MLD for both cases.
    This result validates the mathematical equivalence discussed in \textbf{Remark 3}, confirming that the efficient MLD achieves identical performance to the exhaustive MLD.
    It is also observed that the performance of the approximate MLD closely matches that of the efficient MLD even without prior knowledge of the jamming variance $\sigma_z^2$, which validates the accuracy of the GLRT-based conditional variance estimation in \eqref{eq:var_approx}.
    In both Figs.~\ref{Fig_4}(a) and~\ref{Fig_4}(b), the performance gap between the efficient MLD and the genie-aided MLD with complete knowledge of ${\bf c}$ remains within approximately $0.5$~dB over the entire SNR range.
    This demonstrates that the joint estimation of the jamming indicator vector incurs only a slight performance degradation under strong jamming, which will be further investigated for different SJRs in Fig.~\ref{Fig_7}(a).
    Furthermore, it is observed that the genie-aided upper bound derived in \eqref{eq:BER_upper2} tightly bounds the BER of the genie-aided MLD regardless of the parameters $(p, N, M)$.
    This verifies the accuracy of our theoretical BER analysis presented in Sec.~\ref{Sec:BER}.
    Note that the case with $(p, N, M) = (6, 6, 8)$ in Fig.~\ref{Fig_4}(b) outperforms the case with $(4, 4, 4)$ in Fig.~\ref{Fig_4}(a) due to the higher spreading gain (i.e., $N/S=3$ versus $N/S=2$).

    \begin{figure}[t!]
        \centering
        \includegraphics[width=0.75\columnwidth]{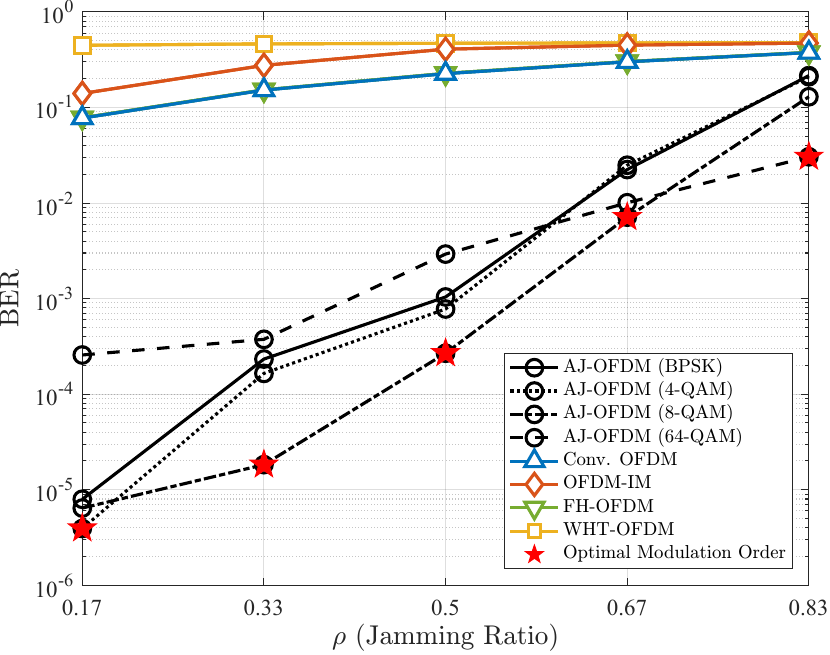}
        \vspace{-2mm}
        \caption{BER comparison of AJ-OFDM with $(p, N) = (6, 6)$ utilizing various modulation orders and existing OFDM schemes versus $\rho$ under partial-band jamming ($\text{SNR} = 20$~dB, $\text{SJR} = -20$~dB). The red stars indicate the BERs of AJ-OFDM with the optimal modulation order.}
        \label{Fig_5}
        \vspace{-3mm}
    \end{figure}

    In Fig.~\ref{Fig_5}, we present the BER comparison of AJ-OFDM with $(p, N) = (6, 6)$ utilizing various modulation orders, and existing OFDM schemes versus the jamming ratio $\rho$ under partial-band jamming at a fixed SNR of $20$~dB and SJR of $-20$~dB.
    As observed, existing OFDM schemes suffer from severe degradation as $\rho$ increases, resulting in an error floor.
    In contrast, AJ-OFDM effectively mitigates the jamming signal even in the high $\rho$ regime, particularly when utilizing a higher modulation order (e.g., 64-QAM).
    This is because a higher $M$ exploits a higher spreading gain of $N/S$, which satisfies the solvability condition, i.e., $N-J \geq S$, while frequency deinterleaving prevents the complete corruption of an individual block.
    Conversely, in the low $\rho$ regime, utilizing a lower modulation order is preferable to increase the minimum Euclidean distance, thereby enhancing robustness against noise.
    Finally, the red stars indicate the BER with the optimal modulation order determined by \eqref{eq:M_opt2}, which accurately tracks the lower envelope of the BER curves.
    These results demonstrate that the proposed scheme achieves superior anti-jamming performance while preserving both the spectral efficiency and average transmit power, at the cost of increased detection complexity, as summarized in Table~\ref{Tab:Scheme}.
    This favorable trade-off is further validated through the simulation results presented in the remainder of this section.

\begin{figure}[t!]
        \vspace{-3mm}
        \centering
        \subfloat[]{\includegraphics[width=0.49\columnwidth]{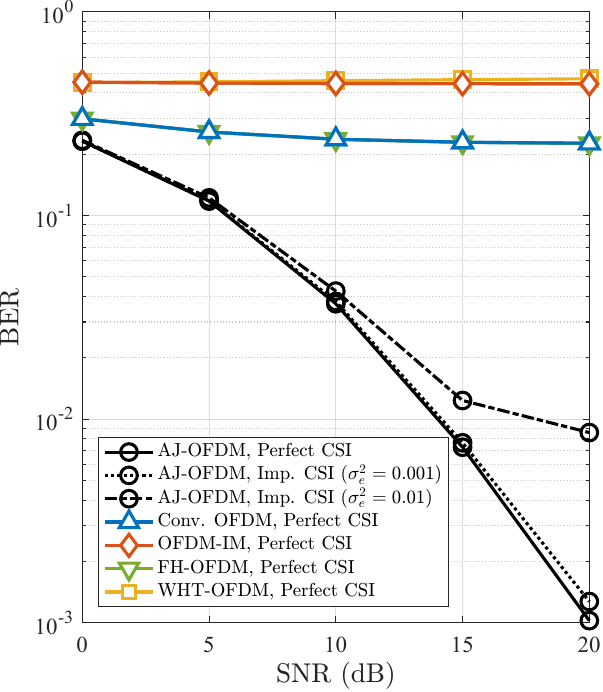}%
        \label{fig:imp_csi}}
        \hfill
        \subfloat[]{\includegraphics[width=0.49\columnwidth]{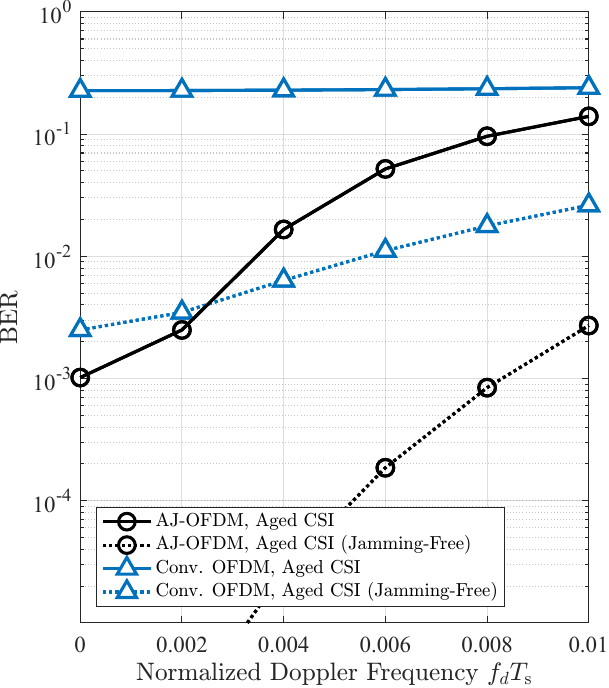}%
        \label{fig:aged_csi}}
        \vspace{-1mm}
        \caption{BER comparison of AJ-OFDM with $(p, N, M) = (4, 4, 4)$ and existing OFDM schemes with channel impairments under partial-band jamming ($\rho=0.5,~\text{SJR}=-20$~dB). (a) BER versus SNR with channel estimation errors. (b) BER versus normalized Doppler frequency $f_d T_{\rm s}$ under a time-varying channel with aged CSI ($\tau=7$, $\text{SNR}=20$~dB).}
        \label{Fig_6}
        \vspace{-3mm}
    \end{figure}

    Fig.~\ref{Fig_6} investigates the BER of AJ-OFDM with $(p, N, M) = (4, 4, 4)$ under channel impairments and partial-band jamming with $\rho=0.5$ and $\text{SJR}=-20$~dB.
    First, Fig.~\ref{Fig_6}(a) compares the BERs of AJ-OFDM and existing OFDM schemes versus SNR with channel estimation errors, i.e., imperfect channel state information (CSI).
    For this analysis, the estimated channel matrix for each block is given by $\hat{\mathbf{H}} = \mathbf{H} + \mathbf{E}$, where $\mathbf{E} = \text{diag}(\bm{\epsilon})$ represents the estimation error matrix with $\bm{\epsilon} \sim \mathcal{CN}(\mathbf{0}_N, \sigma_{\epsilon}^2 \mathbf{I}_N)$.
    The imperfect CSI introduces an error term $-{\mathbf E}{\mathbf U}{\mathbf s}$, which manifests as additional interference, thereby reducing the effective SNR.
    Nevertheless, as shown in Fig.~\ref{Fig_6}(a), AJ-OFDM with imperfect CSI (e.g., $\sigma_{\epsilon}^2 = 0.01$) significantly outperforms existing OFDM schemes even with perfect CSI.
    This is because the spreading gain effectively mitigates the jamming signal, rendering the channel estimation error, which is significantly weaker than the jamming, negligible.

    Next, Fig.~\ref{Fig_6}(b) evaluates the BER of AJ-OFDM under time-varying channels for different normalized Doppler frequencies $f_d T_{\rm s}$, where $f_d$ and $T_{\rm s}$ denote the maximum Doppler frequency and the OFDM symbol duration, respectively.
    Here, $f_d T_{\rm s} = 0.01$ corresponds to a speed of approximately $76$~km/h at a carrier frequency of $2$~GHz with a subcarrier spacing of $15$~kHz, and to higher speeds at lower carrier frequencies, while $f_d T_{\rm s} = 0$ corresponds to the case with perfect CSI.
    For this analysis, the CFR of each subcarrier varies across OFDM symbols according to the Jakes model while remaining i.i.d. across subcarriers, and the aged CSI, i.e., the MMSE prediction of the CFR obtained $\tau = 7$ OFDM symbols before detection, is employed for all schemes.
    Specifically, the MMSE prediction for subcarrier $k$ is given by $J_0(2\pi f_d T_{\rm s} \tau) h_{{\rm f},k}(t-\tau)$, where $J_0(\cdot)$ denotes the zeroth-order Bessel function of the first kind.
    As shown in Fig.~\ref{Fig_6}(b), the BER of AJ-OFDM remains below $3\times10^{-3}$ for $f_d T_{\rm s} \leq 0.002$, which is even lower than that of conventional OFDM without jamming.
    Moreover, AJ-OFDM outperforms conventional OFDM under jamming over the entire range, since the BER of conventional OFDM is dominated by the jammed subcarriers regardless of the channel aging.
    The jamming-free curves exhibit a similar performance degradation, confirming that this degradation is caused by the channel aging rather than the jamming.

    In Fig.~\ref{Fig_7}, we investigate the BERs of AJ-OFDM with $(p, N, M) = (4, 4, 4)$ and existing OFDM schemes under various jamming scenarios and types with $\rho=0.5$ and $\text{SNR}=20$~dB.
    Fig.~\ref{Fig_7}(a) compares the BER versus SJR under partial-band jamming and pulse jamming, where joint time-frequency interleaving with a depth of $T_{\rm int}=28$ is employed for all schemes.
    For pulse jamming, the jammer attacks 14 contiguous OFDM symbols followed by a 14-symbol jamming-free period, maintaining $\rho=0.5$.
    Notably, the performance under pulse jamming is identical to that under partial-band jamming for all schemes, since the joint time-frequency interleaving described in \textbf{Remark~1} disperses burst errors in both the time and frequency domains.
    While existing OFDM schemes suffer from severe degradation in the low SJR regime, AJ-OFDM maintains superior performance across the entire SJR range.
    To further examine the impact of the joint estimation of the jamming indicator vector for different SJRs, Fig.~\ref{Fig_7}(a) also presents AJ-OFDM (Genie), which denotes AJ-OFDM employing the genie-aided MLD.
    It is observed that the BER of AJ-OFDM closely matches that of AJ-OFDM (Genie) in the low SJR regime, which indicates that the jamming indicator vector is accurately estimated under strong jamming.
    Note that a slight performance degradation of AJ-OFDM occurs around $\text{SJR} \approx 0$~dB, as the reliability of the jamming indicator estimation degrades when the jamming and transmitted signal powers are comparable.
    Beyond this regime, the impact of the jamming indicator estimation errors on the BER diminishes with increasing SJR, since the jamming power becomes significantly lower than the transmitted signal power.

    Next, Fig.~\ref{Fig_7}(b) evaluates the robustness of AJ-OFDM and existing OFDM schemes to the reactive jamming discussed in \textbf{Remark~2}, where the jammer is idealized to perfectly observe the transmitted signal and attacks the fraction $\rho = 0.5$ of the subcarriers with the largest transmitted signal amplitudes in each OFDM symbol.
    For each scheme, the jamming signal on the targeted subcarriers follows either the Gaussian distribution or the impulsive distribution, where the impulsive jamming concentrates the jamming power into random spikes with a low activation probability of $0.1$, modeled by the Bernoulli--Gaussian process.
    Both jamming signals are configured to have the same average jamming power $\sigma_z^2$ for a fair comparison, and the approximate MLD derived under the Gaussian assumption is employed for AJ-OFDM without any modification.
    As shown in Fig.~\ref{Fig_7}(b), AJ-OFDM significantly outperforms the existing OFDM schemes under the Gaussian jamming.
    Moreover, even under this idealized observation of the transmitted signal, which is infeasible in practice, AJ-OFDM exhibits only a bounded performance degradation compared with the results in Fig.~\ref{Fig_7}(a), maintaining a BER below $10^{-2}$ over the entire SJR range.
    In contrast, FH-OFDM suffers from severe degradation compared with Fig.~\ref{Fig_7}(a), since the reactive jammer attacks all the activated subcarriers.
    It is worth noting that OFDM-IM achieves a lower BER than in Fig.~\ref{Fig_7}(a), since the jamming signal concentrated only on the activated subcarriers even helps the receiver identify the index bits, and thus the errors are limited to the data symbols on the jammed subcarriers.
    WHT-OFDM exhibits an error floor similar to that in Fig.~\ref{Fig_7}(a), and conventional OFDM is omitted since its identical subcarrier amplitudes make the reactive jamming equivalent to the partial-band jamming in Fig.~\ref{Fig_7}(a).
    Under the impulsive jamming, the BER of AJ-OFDM is even lower than that under the Gaussian jamming, although the approximate MLD is derived under the Gaussian assumption.
    This is because the jamming signal is rarely active and the approximate MLD adapts to the jamming variance of each block through the GLRT-based conditional variance estimation in \eqref{eq:var_approx}.
    These results confirm that the detection performance of the proposed method relies on the magnitude of the residual errors rather than the specific distribution of the jamming signal, as discussed in Sec.~\ref{Sec:ApproxMLD}.

    \begin{figure}[t!]
        \vspace{-3mm}
        \centering
        \subfloat[]{\includegraphics[width=0.49\columnwidth]{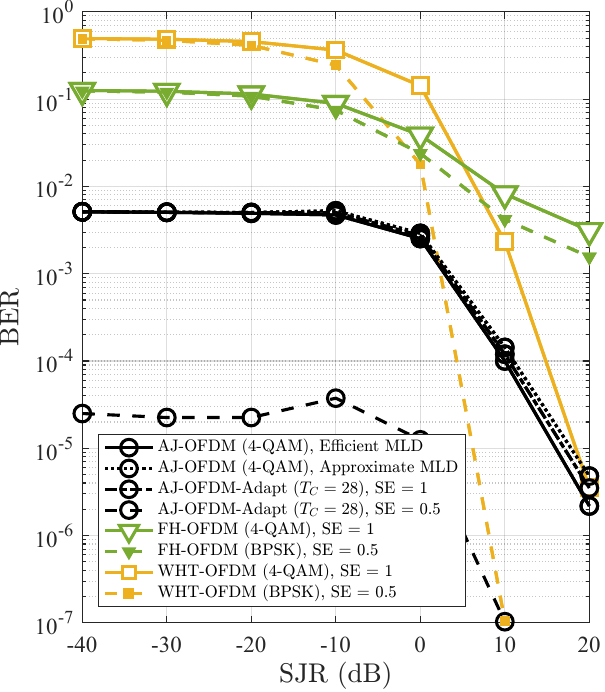}%
        \label{fig:SE}}
        \hfill
        \subfloat[]{\includegraphics[width=0.49\columnwidth]{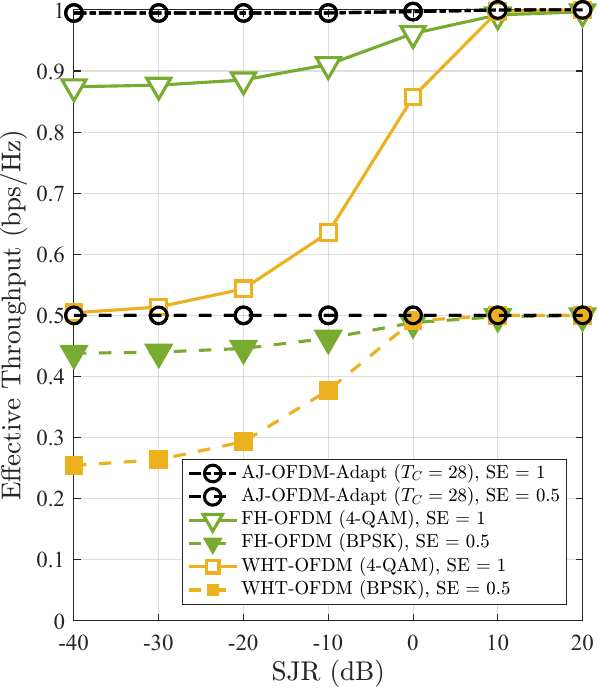}%
        \label{fig:Throughput}}
        \vspace{-1mm}
        \caption{Performance comparison of AJ-OFDM, AJ-OFDM-Adapt, and existing OFDM schemes for different spectral efficiencies (SE) under random jamming ($\rho=0.25,~\text{SNR}=20$~dB). (a) BER versus SJR. (b) Effective throughput versus SJR.}
        \label{Fig_8}
        \vspace{-3mm}
    \end{figure}

    Fig.~\ref{Fig_8} presents the performance comparison of AJ-OFDM, AJ-OFDM-Adapt, and existing OFDM schemes for different spectral efficiencies under random jamming with $\rho=0.25$ and $\text{SNR}=20$~dB.
    For the 1~bps/Hz case ($m=512$), AJ-OFDM utilizes 4-QAM ($p=4,N=4$), while for the 0.5~bps/Hz case ($m=256$), it maintains 4-QAM with ($p=4, N=8$).
    In contrast, FH-OFDM and WHT-OFDM use BPSK for 0.5~bps/Hz.
    As described in Sec.~\ref{Sec:Setup}, all schemes transmit across the same $K$ subcarriers with the same total transmit power, regardless of the spectral efficiency.
    Hence, the different spectral efficiencies are achieved by adjusting only the modulation order or the spreading matrix.
    Fig.~\ref{Fig_8}(a) shows the BER performance versus SJR.
    First, the approximate MLD closely matches the efficient MLD for both spectral efficiencies, which is consistent with the observation in Fig.~\ref{Fig_4}.
    Furthermore, the performance of AJ-OFDM-Adapt ($T_C=28$) lies between that of the efficient MLD and the approximate MLD.
    This is because the adaptive framework employs the approximate MLD during the jamming estimation phase and the efficient MLD during the jamming adaptation phase, resulting in an intermediate performance level.
    In the low SJR regime, AJ-OFDM exhibits superior performance compared to existing schemes.
    Notably, AJ-OFDM at 1~bps/Hz achieves a BER of approximately $5 \times 10^{-3}$ at $\text{SJR}=-20$~dB, whereas the existing OFDM schemes remain above $10^{-1}$ even at a lower spectral efficiency of 0.5~bps/Hz.
    This superiority is achieved because the spreading construction with an adjustable block size enables the joint estimation of the symbol vector and the jamming indicator vector within each block, and thus AJ-OFDM can employ the MLD that achieves optimal detection performance.
    In contrast, WHT-OFDM, which employs the MMSE detector due to the high computational complexity of the MLD for its large block size, suffers from an error floor under severe jamming, since it treats the jamming signal as additional noise without identifying the jammed subcarriers.
    However, in the high SJR regime, the performance of WHT-OFDM improves significantly and approaches that of AJ-OFDM for both 1~bps/Hz and 0.5~bps/Hz.
    This is because, in this regime, the spreading gain of 2 provided by the WHT matrix becomes sufficient to mitigate the jamming signal.
    Next, Fig.~\ref{Fig_8}(b) shows the effective throughput versus SJR, defined as $(p/N) \times (1 - \text{BER})$.
    It is observed that AJ-OFDM-Adapt maintains the target throughput of 1~bps/Hz and 0.5~bps/Hz across the entire SJR range.
    Conversely, the existing schemes suffer from throughput degradation in the low SJR regime, and in particular, the effective throughput of WHT-OFDM at 1~bps/Hz falls to approximately 0.54~bps/Hz at $\text{SJR}=-20$~dB.
    This confirms that the proposed jamming-adaptive communication framework ensures robust and high-rate communication even under severe jamming attacks.
    
    \begin{figure}[t!]
        \centering
        \includegraphics[width=0.75\columnwidth]{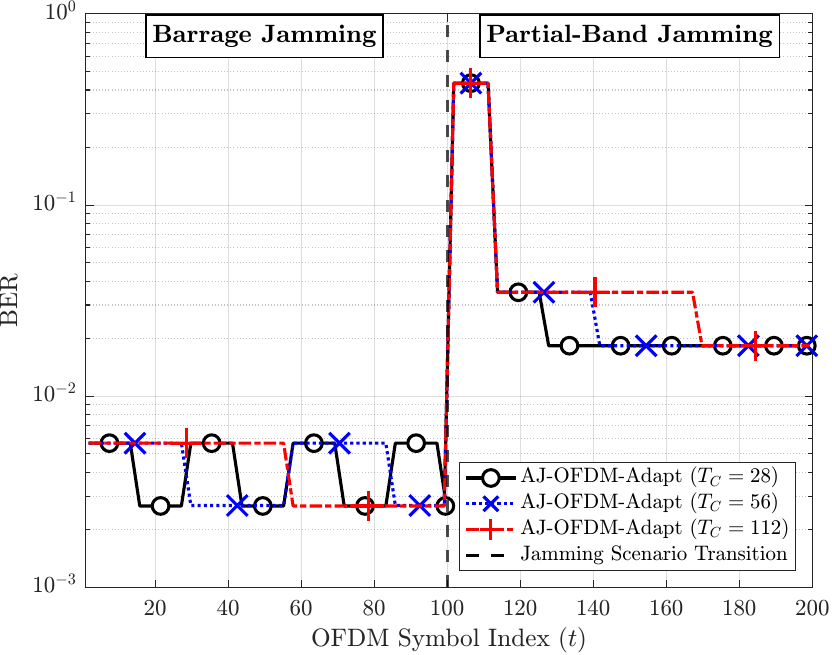}
        \vspace{-2mm}
        \caption{BER comparison of AJ-OFDM-Adapt with $(p, N) = (4, 4)$ for different $T_C$ versus OFDM symbol index under a dynamic jamming scenario. The jamming scenario transitions from barrage jamming ($\rho=1,~\text{SJR}=10$~dB) to partial-band jamming ($\rho=0.75, \text{SJR}=-20$~dB) at $t=100$.}
        \label{Fig_9}
        \vspace{-3mm}
    \end{figure}   

    In Fig.~\ref{Fig_9}, we present the BER comparison of AJ-OFDM-Adapt with $(p, N) = (4, 4)$ for different transmission cycles $T_C$ versus the OFDM symbol index under a dynamic jamming scenario.
    The simulation starts under barrage jamming with $\rho=1$ and $\text{SJR}=10$~dB, and transitions to partial-band jamming with $\rho=0.75$ and $\text{SJR}=-20$~dB at the 100th OFDM symbol.
    In the first jamming scenario ($t < 100$), the BER exhibits a step-like behavior.
    The higher BER periods correspond to the jamming estimation phase employing the approximate MLD, while the lower BER periods correspond to the adaptation phase employing the efficient MLD.
    It is worth noting that under barrage jamming, even though all subcarriers are corrupted, the MLD remains the optimal detection scheme as it effectively aggregates the signal energy spread across all subcarriers.
    Thanks to this capability, AJ-OFDM-Adapt maintains robust performance, benefiting from the spreading gain and frequency diversity gain.
    At $t=100$, the transition to severe partial-band jamming causes a sharp increase in BER, as the modulation order optimized for the previous scenario becomes unsuitable for the new environment.
    However, AJ-OFDM-Adapt adaptively recovers the performance by updating the modulation order in the subsequent transmission cycle based on the new jamming parameters.
    As shown in Fig.~\ref{Fig_9}, while $T_C=28$ adapts most rapidly to the change, $T_C=112$ exhibits a delayed response, maintaining a high BER longer before convergence.
    Although all cases eventually converge to the same performance level, this result highlights the trade-off between the adaptation speed and the feedback overhead.
    Therefore, the transmission cycle $T_C$ should be selected considering the dynamics of the jamming environment.

    \section{Conclusion}

    In this paper, we have proposed a novel anti-jamming OFDM scheme to ensure robust communication even under severe jamming attacks while maintaining high spectral efficiency.
    Specifically, we have designed an anti-jamming modulation method that transforms the symbol vector into a higher-dimensional modulated vector using a spreading matrix, thereby exploiting both the spreading gain and the frequency diversity gain.
    To recover the symbols efficiently, we have developed the efficient MLD method and proved its mathematical equivalence to the exhaustive MLD method via the rearrangement inequality.
    Furthermore, based on the derived theoretical BER upper bound, we have formulated an optimization problem to determine the optimal modulation order according to the jamming environment.
    To ensure practicality without prior knowledge of jamming parameters, we have established a jamming-adaptive communication framework that employs the GLRT principle to estimate jamming parameters and adapt to dynamic jamming scenarios.
    Simulation results have demonstrated that the proposed scheme significantly outperforms existing OFDM schemes in both BER and effective throughput under various jamming scenarios.
    Interesting avenues for future research include extending the framework to counter passive jamming attacks, such as DISCO jamming \cite{Disco}, and incorporating jamming pattern prediction.

    \bibliographystyle{IEEEtran}
    \bibliography{Reference}
\end{document}